\documentclass[11pt]{article}
\pdfoutput=1

\usepackage[a4paper,top=3cm,bottom=4cm]{geometry}

\newcommand*{\changed}{}
\usepackage{Packages}

\usepackage{Definitions}

\newcommand{\OfficialTitle}{Relating Gauge Theories via\\ Gauge/Bethe Correspondence}

\hypersetup{pdfauthor={Domenico Orlando and Susanne Reffert},pdftitle={\OfficialTitle}}

\author{
  \begin{minipage}{.97\linewidth}
    \vspace{1cm}
    \begin{center}
      \begin{small}
        \textbf{Domenico Orlando} and \textbf{Susanne Reffert}
      \end{small}
    \end{center}
      \vspace{1cm} \hspace{4cm}\begin{minipage}{.7\linewidth}
        {\it \begin{footnotesize}
            Institute for the Mathematics and Physics of
            the Universe, \\The University of Tokyo, Kashiwa-no-Ha
            5-1-5, \\ Kashiwa-shi, 277-8568 Chiba, Japan. \\
          \end{footnotesize}}
      \end{minipage}
      \vspace{1cm}
    \end{minipage}
}

\date{}

\title{\vspace{3cm}
  \begin{huge}
    \textbf{\OfficialTitle}
  \end{huge}
}

\begin{document}

\setstretch{1.1}

\numberwithin{equation}{section}

\begin{titlepage}
  \maketitle
  \thispagestyle{empty}

  \vspace{-14cm}
  \begin{flushright}
    IPMU10-0088
  \end{flushright}

  \vspace{14cm}

  \begin{center}
    \textsc{Abstract}
  \end{center}
  In this note, we use techniques from integrable systems to study
  relations between gauge theories.  The Gauge/Bethe correspondence,
  introduced by Nekrasov and Shatashvili, identifies the
  supersymmetric ground states of an $\mathcal{N}=(2,2)$
  supersymmetric gauge theory in two dimensions with the Bethe states
  of a quantum integrable system.  We make use of this correspondence
  to relate \changed different quiver gauge theories which correspond to
  different formulations of the Bethe equations of the \textsc{xxx}
  and 
  the $tJ$ models.

\end{titlepage}

\tableofcontents

\section{Introduction}
\label{sec:introduction}

In this note, we use techniques from integrable systems to study
relations between gauge theories.  There have been many examples of
the interplay between gauge theories (often with string theory
interpretations) and integrable
systems\changed~\cite{Martinec:1995by,Donagi:1995cf,Gorsky:1996hs,Gorsky:1997jq,Gorsky:1997mw,Boels:2004ua,Faddeev:1994zg,Minahan:2002ve,Beisert:2006ez,Dijkgraaf:2007yr,Dijkgraaf:2008ua,Dijkgraaf:2009gr}.
In~\cite{Nekrasov:2009uh, Nekrasov:2009ui}, a beautiful dictionary
between quantum integrable systems such as spin chains, and
$\mathcal{N}=(2,2)$ supersymmetric gauge theories in two dimensions
was introduced. In short, the supersymmetric \emph{ground states} of
the gauge theory are mapped directly to the \emph{Bethe spectrum} of
the integrable model. The \emph{effective twisted superpotential} of
the gauge theory in the Coulomb branch is identified with the
\emph{Yang--Yang counting function} which serves as a potential for
the Bethe equations, whose solutions are the spectrum of the
integrable system.  In~\cite{Nekrasov:2009rc}, this correspondence was
extended to four dimensional gauge theories which correspond to Toda
or Calogero--Moser models. A brane construction in the topological
string A--model for the
theories described in~\cite{Nekrasov:2009rc} was provided in~\cite{Nekrasov:2010ka}.  The Gauge/Bethe
correspondence is thought to encompass also the
AGT--correspondence~\cite{Alday:2009aq} which was explained from
the point of view of matrix models in~\cite{Dijkgraaf:2009pc}.

\bigskip

Here we will concentrate on the simpler correspondence between spin
chain--type integrable models and supersymmetric gauge theories in two
dimensions. We use the Gauge/Bethe correspondence to make a statement
about the supersymmetric ground states of seemingly unrelated gauge
theories.  We use the fact that integrable models 
can \changed give rise to several different systems of
Bethe equations, which nonetheless lead to the same spectrum. Since
the Gauge/Bethe correspondence relates the Yang--Yang function to the
effective twisted superpotential, different Bethe equations correspond
to different gauge theories. This means that the Gauge/Bethe
correspondence can be used to relate the low energy properties of
different gauge theories. By the correspondence, we know that gauge
theories which can be traced back to equivalent Bethe equations have
the same supersymmetric ground states.\\
\changed The simplest instance of this phenomenon is found
in the \textsc{xxx} spin chain with twisted boundary conditions, where
configurations with $N$ magnons can be mapped into configurations with
$L - N$ magnons which describe the same state. Another, richer,
example is the so--called $tJ$ \emph{model}. The fact that the $tJ$
spin chain has the supergroup $sl(1|2)$ as symmetry group leads to
three different, but ultimately equivalent sets of Bethe equations,
corresponding to the different choices of the Cartan matrix of
$sl(1|2)$. On the other side of the correspondence, we have three
quiver--type gauge theories with two different gauge groups
each. Their supersymmetric ground states are thus shown to be
equivalent.  Note that our statement is stronger than a mere counting
of the vacua of the three theories. The three sets of Bethe equations
are actually different ways of writing the same non--linear
conditions. It is reasonable to expect the correspondence to go beyond
the ground states and to relate also the solitons which interpolate
between them.  To give more weight to our claim, we show that at least
in the case of zero twisted masses, it is possible to embed the three
quiver gauge theories into string theory in terms of brane cartoons of
\D2, \D4 and \NS5--branes, and that it is possible to relate them via
brane motions.
While we exemplify our approach using the $tJ$ model, we believe that
it can be used in a wider context.  We believe that new insights into
supersymmetric gauge theories in two dimensions can be obtained by
expressing them as quantum integrable systems.


The plan of this note is the following. In
Section~\ref{sec:n=2-supersymm-gauge}, we briefly review
$\mathcal{N}=(2,2)$ supersymmetric gauge theories in two dimensions.
We introduce the field content and action (\S~\ref{sec:notation}), the
low energy effective action (\S~\ref{sec:low-energy}), and finally the
three quiver gauge theories we set out to relate in this note
(\S~\ref{sec:examples}).  In Section~\ref{sec:correspondence}, we
introduce integrable spin chains (\S~\ref{sec:parameters}), spell out
the necessary knowledge of the algebraic Bethe ansatz
(\S~\ref{sec:algebr-bethe-ansatz}), state the Gauge/Bethe
correspondence as described by Nekrasov and Shatashvili
(\S~\ref{sec:dictionary}), and introduce the $tJ$ model
(\S~\ref{sec:example:-tj-model}). The explicit matching is done in
\S~\ref{sec:corrtJ}.  As an alternative justification, brane cartoons
for the massless cases of our quiver gauge theories are introduced in
Section~\ref{sec:brane-cartoons}, where also their relation via brane
transitions is detailed. In Section~\ref{sec:conclusions}, we conclude
with a view on more general applications of the approach taken in this
note. The basics of the superalgebra $sl(1|2)$ are collected in
Appendix~\ref{sec:superalgebra}, while the proof of the equivalence of
the three Bethe equations underlying the gauge theories is given in
Appendix~\ref{sec:equivalence}.


\section{${\mathcal N}=(2,2)$ supersymmetric gauge theories in two dimensions}
\label{sec:n=2-supersymm-gauge}

In this section, we will introduce the necessary notation for dealing
with $\mathcal{N}=(2,2)$ gauge theories in two dimensions, discuss
their low energy action after integrating out massive fields, and
finally present three seemingly unrelated quiver gauge theories, which
will be our explicit examples throughout this note.

\subsection{Field content and action}
\label{sec:notation}

Let us quickly review the basics of $\mathcal{N}=(2,2)$ theories. We
will use the notation of~\cite{Hori:2003ic}; for greater detail, we
refer our readers to~\cite{Hori:2003ic,Witten:1993yc,Hanany:1997vm}.

$\mathcal{N}=(2,2)$ theories are field theories in $1+1$ dimensions with
two (real) positive and two (real) negative chirality
supercharges. Superspace is described by the two bosonic coordinates
$x^0,\ x^1,$ and the four fermionic coordinates $\theta^+,\ \theta^-,\
\bar\theta^+,\ \bar\theta^-$.  We define the differential operators
\begin{equation}
  D_{\pm}=\frac{\partial}{\partial\theta^\pm}-i\bar\theta^\pm\partial_\pm,\quad \overline D_{\pm}=-\frac{\partial}{\partial\bar\theta^\pm}+i\bar\theta^\pm\partial_\pm.
\end{equation}
The $\theta$--expansion of the \emph{vector} superfield in
Wess--Zumino gauge is given by
\begin{multline}
  V= \theta^-\bar\theta^-(A_0-A_1)+\theta^+\bar\theta^+(A_0+A_1)-\theta^-\bar\theta^+\sigma-\theta^+\bar\theta^-\overline\sigma\\
  +i\theta^-\theta^+(\bar\theta^-\overline\lambda_-+\bar\theta^+\overline\lambda_+)+i\bar\theta^+\bar\theta^-(\theta^-\lambda_-+\theta^+\lambda_+)+\theta^-\theta^+\bar\theta^+\bar\theta^-D,
\end{multline}
where $A_\mu$ is a vector field, $\lambda_\pm,\ \overline\lambda_\pm$ are Dirac fermions which are conjugate to each other, $\sigma$ is a complex scalar, and $D$ is a real auxiliary field. With this, we can now define the gauge covariant derivative
\begin{equation}
\mathcal{D}_\pm=e^{-V}D_\pm e^V, \quad \quad \overline{\mathcal{D}}_\pm=e^{V}\overline D_\pm e^{-V}.
\end{equation}
A \emph{chiral} superfield satisfies $\overline{\mathcal{D}}_\pm\Phi=0$. The $\theta$--expansion of the chiral superfield is given by
\begin{equation}
  \Phi=\phi(y^\pm)+\theta^\alpha\psi_\alpha(y^\pm)+\theta^+\theta^-F(y^\pm),
\end{equation}
where $\phi$ is a complex scalar field, $\psi_\alpha$ a Dirac fermion, $F$ a complex auxiliary field, $y^\pm=x^\pm-i\theta^\pm\bar\theta^\pm$, and $x^\pm=x_0\pm x^1$. 
A \emph{twisted chiral} superfield satisfies $\overline{\mathcal{D}}_+\Sigma=\mathcal{D}_-\Sigma=0$.
 The super field strength $\Sigma=\tfrac{1}{2}\{\overline{\mathcal{D}}_+,\mathcal{D}_-\}$ is a twisted chiral superfield and its $\theta$--expansion is given by
\begin{equation}
  \Sigma=\sigma(\tilde y^\pm)+i\theta^+\overline\lambda_+(\tilde y^\pm)-i\bar\theta^-\lambda_-(\tilde y^\pm)+\theta^+\bar\theta^-[D(\tilde y^\pm)-iA_{01}(\tilde y^\pm)] + \dots \, ,
\end{equation}
where $\tilde y^\pm=x^\pm\mp i\theta^\pm\bar\theta^\pm$, and $A_{01}=\partial_0A_1-\partial_1A_0+[A_0,A_1]$.

The $\mathcal{N}=(2,2)$ field theory in two dimensions can be understood
as the dimensional reduction of $\mathcal{N}=1$ supersymmetric gauge
theory in four dimensions. The scalar $\sigma$ results from the
$x^{2,3}$ components of the vector field $A_\mu$ in four
dimensions.\footnote{It is possible to obtain the same theory by
  dimensional reduction from six dimensions. In this case, the
  $x^{4,5}$ components of the vector field turn into the complex
  scalar $\phi$ of the chiral multiplet in the adjoint
  representation. } \bigskip

In the supersymmetric action, there are three kinds of couplings:
\begin{itemize}
\item the \emph{$D$--term}: $\int \di^2 x\, \di^4\theta \, K$,
  where $K$ is an arbitrary (real) differential function of the
  superfields,
\item the \emph{$F$--term} (plus its Hermitian conjugate) :
  $\int\di^2x\,\di\theta^- \di \theta^+  \left. W
  \right|_{\bar\theta_\pm=0}\ + \text{h.c.}$, where the
  \emph{superpotential} $W$ is a holomorphic function of the chiral
  multiplets,
\item the \emph{twisted $F$--term} (plus its Hermitian conjugate):
  $\int \di^2x\,\di\bar\theta^- \di\theta^+
  \left. \widetilde W \right|_{\bar\theta_+=\theta^-=0} + \text{h.c.}$, where
  the \emph{twisted superpotential} $\widetilde W$ is a holomorphic
  function of the twisted superfields.
\end{itemize}
The supersymmetric structure implies some decoupling and
non--renormalization theorems: in particular neither the $F$--term nor
the twisted $F$--term get renormalized. Moreover, in the effective
action, the $F$--term and twisted $F$--term cannot mix.

Let us consider a gauge theory with gauge group $G$ and chiral matter
multiplets $X_k$ (we denote them by $Q$ if they are in the
fundamental, $\overline Q$ if they are in the anti--fundamental, $B$
if they are in the bifundamental, and $\Phi$ if they are in the
adjoint representation). The kinetic term of the Lagrangian is given
by
\begin{equation}
  L_{\text{kin}}=\int\di^4\theta \left(\sum_k X_k^\dagger\, e^V X_k - \frac{1}{2e^2}\Tr (\Sigma^\dagger\Sigma)\right),
\end{equation}
where $e$ is the gauge field strength. We can consider the following
additional terms:
\begin{itemize}
\item {\em Fayet--Iliopoulos} (FI) and theta--term:
  $L_{\text{FI},\vartheta}=-\tfrac{\imath}{2}\tau\int \di\bar\theta^-
  \di\theta^+ \Tr\Sigma + \text{ h.c.\,}$, where
  $\tau=ir+\vartheta/2\pi$. Such a term can be turned on for every $U(1)$--factor in the center of $G$.
\item The \emph{complex mass}: $L_{\text{mass}}=\displaystyle{\sum_{k,l}} \int
  \di^2 \theta\, m_k^{\phantom{k}l} \widetilde X_l X^k + \text{h.c.}$,
  where $m_k^{\phantom{k}l}$ are complex parameters.
\item The \emph{twisted masses}: $L_{\text{tw}}=\int \di^4\theta \,
  (X^\dagger e^{\theta^-\bar\theta^+\widetilde m_X + \text{h.c.}}X)$,
  where $e^{\theta^-\bar\theta^+\widetilde m_X}$ are matrices in the
  same representation as $X$ of the maximal torus of the global
  symmetry group\footnote{In the rest of this note, all the twisted
    masses are defined up to a scale factor $u$ that we set equal to
    $1$.}.
\end{itemize}

While the complex masses are already present in the four dimensional
theory, this is not the case for the twisted masses. The twisted
masses are deformations of the theory which are related to the global
symmetry group of the theory. They can be obtained by first gauging
the global symmetry group, giving a vev to the scalar component of the
vector superfield, and in the end making the fields
vanish~\cite{Hanany:1997vm}.  The global symmetry group $H$ of an
$\mathcal{N}=(2,2)$ gauge theory receives a factor of $U(L_i)$ for
each set of $L_i$ fundamental or anti--fundamental multiplets
$Q^i$. For each adjoint multiplet $\Phi^i$ and each bifundamental
multiplet $B^{ij}$, $H$ receives a further factor of $U(1)$.
Moreover, there are two $U(1)$ $R$--symmetries which are usually
denoted by $U(1)_V$ and $U(1)_A$.  Turning on twisted masses will
break $H$ down to its maximal torus.  Also turning on a superpotential
will in general break the global symmetry group to some extent.  This
implies that a general superpotential will be incompatible with
general twisted masses. However, special choices of the superpotential
and twisted mass parameters are possible which allow both deformations
to coexist.

Once the twisted masses are turned on, the matter fields become
massive and can be integrated out to obtain a low energy
effective action.

\subsection{Low energy effective action}
\label{sec:low-energy}

In this section, we describe the \emph{Coulomb branch} of the theory. We
therefore consider the low energy effective theory obtained for slowly
varying $\sigma$ fields after integrating out the massive matter
fields. In this way, we obtain an effective twisted superpotential
$\widetilde W_{\text{eff}} (\Sigma) $; the vacua of the
theory~\cite{Nekrasov:2010ka} are the solutions of the equation
\begin{equation}
\label{eq:vacua}
  \exp \left[ 2\pi \frac{\partial \widetilde W_{\text{eff}}(\sigma)}{\partial \sigma_i} \right]  = 1 \, .  
\end{equation}

Consider the  case of a $U(1)$ gauge theory with one chiral superfield
$Q$ of charge~$1$ and twisted mass $\widetilde m_Q$. The most general
supersymmetric action containing terms with at most four fermions and
two derivatives is given by
\begin{equation}
  S_{\text{eff}}(\Sigma) = - \int \di^4 \theta \, K_{\text{eff}} ( \Sigma, \overline{\Sigma} ) + \frac{1}{2} \int \di^2 \theta \, \widetilde W_{\text{eff}} (\Sigma) + \text{h.c.}  \,.
\end{equation}
In the absence of an $F$--term, the action $S(\Sigma, Q)$ is quadratic in
$Q$, and the effective action can be evaluated exactly via a one--loop
calculation:
\begin{equation}
  e^{\imath S_{\text{eff}}(\Sigma)} = \int \mathcal{D} Q \, e^{\imath S(\Sigma, Q)} \, .  
\end{equation}
The bosonic determinant equals
\begin{equation}
  \int \frac{\di^2 k }{\left( 2 \pi \right)^2} \log ( k^2 + \abs{\sigma - \widetilde m_Q}^2 + D ) \, ,  
\end{equation}
and expanding in powers of $D$,
\begin{equation}
  \log ( k^2 + \abs{\sigma - \widetilde m_Q}^2 + D )  = \log ( k^2 + \abs{\sigma - \widetilde m_Q}^2  ) + \frac{D}{k^2 + \abs{\sigma - \widetilde m_Q }^2} + \dots
\end{equation}
The zeroth order term is cancelled by the fermionic determinant, while
the first order term leads, after integrating over the momenta,
to the \emph{effective twisted superpotential}
\begin{equation}
  \widetilde W_{\text{eff}} (\Sigma) = \frac{1}{2\pi}\left( \Sigma - \widetilde m_Q \right) \left( \log (\Sigma - \widetilde m_Q) - 1 \right) - \imath \tau \,\Sigma \, , 
\end{equation}
where we also added the contribution of the Fayet--Iliopoulos term.
In the general case, an $F$--term is possible but, thanks to the
\emph{decoupling theorem}, it would not change the expression of the
effective twisted superpotential.

In the following, we will be mainly interested in \emph{quiver gauge
theories}. In this case, the gauge group $G$ and the flavor group $F$ are
direct products:
\begin{align}
  G &= \prod_{a=1}^r U(N_a) \, ,& F &= \prod_{a=1}^r U(L_a) \, .
\end{align}
These theories can be represented via \emph{quiver diagrams}. Each
factor $U(N_a)$ corresponds to a node, a bifundamental field in the
representation $\overline{\mathbf{N_a}} \otimes \mathbf{N_b}$ is
denoted by an arrow going from node $a $ to node $b$, and an adjoint
field is an arrow starting and ending on the same node. Each component
$U(L_a)$ of the flavor group is represented by an extra node, joined
by a dotted arrow to the relevant component of the gauge group (see
Figure~\ref{fig:quivers} for some examples).  The evaluation of the
effective twisted superpotential in the non--Abelian case is very
similar to the $U(1)$ calculation once one observes that the classical
vacuum equations require $\sigma$ to be diagonalizable. If we assume
that $\sigma_i \neq \sigma_j $ for $i \neq j$, which breaks the gauge
group $U(N)$ to its maximal torus $U(1)^N$, we can perform exactly the
same Gaussian integration of the chiral fields as above. We thus
obtain the following contributions to the effective twisted
superpotential:
\begin{itemize}
\item For each fundamental field $Q_k$ with twisted mass $\widetilde m_k^\fund $:
  \begin{equation}
    \widetilde W_{\eff}^\fund = \frac{1}{2\pi}\sum_{i=1}^N \left( \sigma_i - \widetilde m^\fund_k \right) \left( \log ( \sigma_i - \widetilde m^\fund_k ) - 1 \right).
  \end{equation}
\item For each anti--fundamental field $\overline{Q}_k $ with twisted mass $\widetilde m_k^{\bar \fund}$:
  \begin{equation}
    \widetilde W_{\eff}^{\bar \fund} = \frac{1}{2\pi}\sum_{i=1}^N \left( - \sigma_i - \widetilde m^{\bar \fund}_k \right) \left( \log ( - \sigma_i - \widetilde m^{\bar \fund}_k ) - 1 \right).
  \end{equation}
\item For each adjoint field $\Phi$ with twisted mass $\widetilde m^{\adj}$:
  \begin{equation}
    \widetilde W_\eff^\adj = \frac{1}{2\pi}\sum_{\substack{i,j=1\\i \neq j}}^{N} \left( \sigma_i - \sigma_j - \widetilde m^\adj \right) \left( \log (\sigma_i - \sigma_j - \widetilde m^\adj ) - 1 \right).     
  \end{equation}
\item For each bifundamental $B^{12}$ in the representation
  $\overline{\mathbf{N_1}} \otimes \mathbf{N_2} $ and twisted mass $\widetilde m^\bif$:
  \begin{equation}
    \widetilde W_\eff^\bif = \frac{1}{2\pi}\sum_{i=1}^{N_1} \sum_{p=1}^{N_2} \left( - \sigma^{(1)}_i + \sigma^{(2)}_p - \widetilde m^\bif \right) \left( \log ( - \sigma^{(1)}_i + \sigma^{(2)}_p - \widetilde m^\bif ) - 1 \right) ,    
  \end{equation}
  where the $\sigma^{(a)}_i$ are the scalar components of the vector multiplet for the group $U(N_a)$.
\end{itemize}

\subsection{Example: two gauge theories}
\label{sec:example:-}

\changed Consider the $\mathcal{N}=(2,2)$ theory with gauge group $U(N)$ and
the following matter content~\cite{Nekrasov:2009uh}:
\begin{itemize}
\item an adjoint field $\Phi$ with twisted mass $\imath$,
\item $L$ fundamentals and anti--fundamentals $Q_k, \overline{Q}_k$
  with twisted mass $-\imath /2 $.
\end{itemize}
The effective twisted superpotential is given by:
\begin{multline}
  \label{eq:XXX-1}
  \widetilde W^N_{\text{eff}} ( \sigma ) = \frac{L}{2\pi} \sum_{i=1}^{N}
  \left[ \left( \sigma_i + \frac{\imath}{2} \right) \left( \log
      (\sigma_i + \frac{\imath}{2}) - 1 \right) - \left( \sigma_i -
      \frac{\imath}{2} \right) \left( \log (-
      \sigma_i + \frac{\imath}{2}) - 1 \right) \right] \\
  + \frac{1}{2\pi} \sum_{\substack{i,j\\i \neq j }}^{N} \left(
    \sigma_i - \sigma_j - \imath \right) \left( \log (\sigma_i -
    \sigma_j - \imath ) - 1 \right) - \imath \tau \sum_{i=1}^{N}
  \sigma_i \, .
\end{multline}

We intend to compare it to a system with gauge group $U(L-N)$, $L$
fundamental and antifundamentals and opposite Fayet--Iliopoulos term,
which admits the following effective twisted superpotential:
\begin{multline}
  \label{eq:XXX-2}
  \widetilde W_{\text{eff}}^{L-N} ( \sigma ) = \frac{L}{2\pi} \sum_{i=1}^{L-N}
  \left[ \left( \sigma_i + \frac{\imath}{2} \right) \left( \log
      (\sigma_i + \frac{\imath}{2}) - 1 \right) - \left( \sigma_i -
      \frac{\imath}{2} \right) \left( \log (-
      \sigma_i + \frac{\imath}{2}) - 1 \right) \right] \\
  + \frac{1}{2\pi} \sum_{\substack{i,j\\i \neq j }}^{L-N} \left(
    \sigma_i - \sigma_j - \imath \right) \left( \log (\sigma_i -
    \sigma_j - \imath ) - 1 \right) + \imath \tau \sum_{i=1}^{N}
  \sigma_i \, .
\end{multline}

\subsection{Example: three quiver gauge theories}
\label{sec:examples}

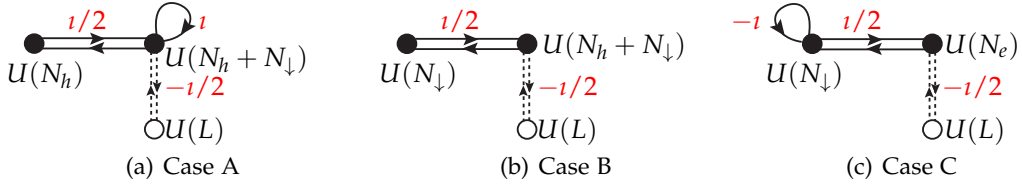
\begin{figure}
  \centering
  \subfigure[Case A]{ 
    \label{fig:quiver-A}
    \begin{picture}(130,45)(,)
      \begin{small}
        \put(0,22){$U(N_h)$} \put(59,28){$U(N_h + N_\da) $} \put(59,1){$U(L)$}
        \textcolor{red}{\put(72,42){$\imath$} \put(23,42){$\imath/2$} \put(59,17){$-\imath/2$}}
      \end{small}
      \includegraphics[scale=.7]{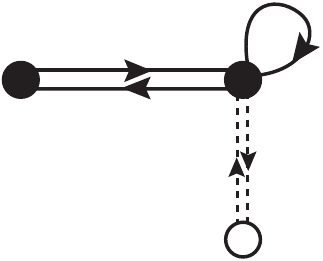}
    \end{picture} }   
  \subfigure[Case B]{ 
    \label{fig:quiver-B}
    \begin{picture}(130,45)(,)
      \begin{small}
        \put(0,22){$U(N_\da)$} \put(61,33){$U(N_h + N_\da) $} \put(59,1){$U(L)$}
         \textcolor{red}{\put(23,42){$\imath/2$} \put(59,17){$-\imath/2$}}
      \end{small}
      \includegraphics[scale=.7]{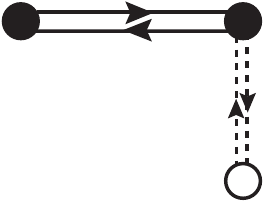}
    \end{picture} }   
  \subfigure[Case C]{ 
    \label{fig:quiver-C}
    \begin{picture}(110,45)(,)
      \begin{small}
        \put(5,22){$U(N_\da)$} \put(72,33){$U(N_e)$} \put(72,1){$U(L)$}
        \textcolor{red}{\put(-8,42){$-\imath$} \put(34,42){$\imath/2$} \put(71,17){$-\imath/2$}}
      \end{small}
      \includegraphics[scale=.7]{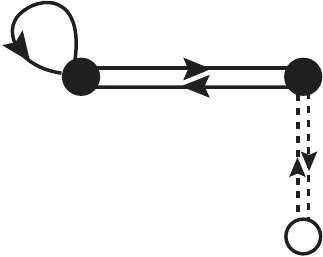}
    \end{picture} }   
  \caption{Quiver diagrams for the three example theories. The
    twisted masses for the chiral fields are given in red.}
  \label{fig:quivers}
\end{figure}

The main claim of this note is that the three quiver gauge theories
described below, which have different gauge groups and different
matter contents, share the same chiral ring (and therefore have the
same supersymmetric vacua). The three theories are given as follows:

\paragraph{Case A} A quiver gauge theory with gauge groups $U(N_h +
N_\da)$ and $U(N_h)$ (the reason for the names of the parameters $N_h$ and
$N_\da$ will become clear in the following), with the following matter content:
\begin{itemize}
\item a bifundamental $B^{12}$ in the representation $
  \overline{\mathbf{N_h}} \otimes (\mathbf{N_h + N_\da} )$ with
  twisted mass $\imath/2$,
\item a bifundamental  $B^{21}$ in the representation
  $(\overline{\mathbf{N_h + N_\da}})\otimes \mathbf{N_h}$ with twisted
  mass $\imath/2$,
\item an adjoint  $\Phi^2$ for $U(N_h + N_\da)$ with twisted mass
  $\imath$,
\item $L$ fundamentals and anti--fundamentals $(Q^2_{\phantom{2}k},
  \overline{Q^2}_k)$ for $U(N_h+N_\da)$ with twisted mass $-\imath /2
  $.
\end{itemize}
The global symmetry group $H$ (which is broken down to its maximal
torus by the twisted masses) is $U(L)_Q\times U(L)_{\widetilde
  Q}\times U(1)_B\times U(1)_{\widetilde B}\times U(1)_\Phi$. The
quiver diagram is represented in Figure~\ref{fig:quiver-A}. Using the
results above, we find the following effective twisted superpotential 
\begin{small}
  \begin{multline}
    \label{eq:Weff-A}
    \widetilde W_{\text{eff}}^A ( \sigma ) = \frac{L}{2\pi}
    \sum_{p=1}^{N_h + N_\da} \left[ \left( \sigma_p^{(2)} +
        \frac{\imath}{2} \right) \left( \log (\sigma_p^{(2)} +
        \frac{\imath}{2}) - 1 \right) - \left( \sigma_p^{(2)} -
        \frac{\imath}{2} \right) \left( \log (-
        \sigma_p^{(2)} + \frac{\imath}{2}) - 1 \right) \right] \\
    + \frac{1}{2 \pi}\sum_{i=1}^{N_h} \sum_{p=1}^{N_h + N_\da} \left[
      \left( \sigma_i^{(1)} - \sigma_p^{(2)} - \frac{\imath}{2}
      \right) \left( \log ( \sigma_i^{(1)} - \sigma_p^{(2)} -
        \frac{\imath}{2} ) - 1 \right) \right. \\ \left. - \left(
        \sigma_i^{(1)} - \sigma_p^{(2)} + \frac{\imath}{2} \right)
      \left( \log ( - \sigma_i^{(1)} + \sigma_p^{(2)} -
        \frac{\imath}{2} ) - 1 \right) \right] \\ + \frac{1}{2\pi}
    \sum_{\substack{p,q\\p \neq q }}^{N_h+N_\da} \left( \sigma_p^{(2)}
      - \sigma_q^{(2)} - \imath \right) \left( \log (\sigma_p^{(2)} -
      \sigma_q^{(2)} - \imath ) - 1 \right) - \imath \tau_1
    \sum_{i=1}^{N_h} \sigma^{(1)}_i - \imath \tau_2
    \sum_{p=1}^{N_h+N_\da} \sigma^{(2)}_p \, .
  \end{multline}
\end{small}%
The twisted masses are compatible with a superpotential of the type
\begin{equation}
  W_A (Q^2,\overline{Q^2}, \Phi^2, B^{12}, B^{21}) = \sum_{k} \left[ a \, Q^2_{\phantom{2}k} \Phi^2 \overline{Q^2}_k + b \, Q^2_{\phantom{2}k} B^{21}  B^{12} \overline{Q^2}_k \right] \, ,
\end{equation}
where $a$ and $b$ are parameters.

\paragraph{Case B} A quiver gauge theory with gauge groups $U(N_h +
N_\da)$ and $U(N_\da)$, with the following matter content:
\begin{itemize}
\item  a bifundamental $B^{12}$ in the representation
$\overline{\mathbf{N_\da}} \otimes (\mathbf{N_h+ N_\da})$ and twisted
mass $\imath/2$, 
\item a bifundamental $B^{21}$ in the representation $(
\overline{\mathbf{N_h + N_\da}} )\otimes \mathbf{N_\da}$ and twisted
mass $\imath/2$,
\item $L$ fundamentals and anti--fundamentals $(Q^2_{\phantom{2}k},
\overline{Q^2}_k)$ for $U(N_h + N_\da)$ with twisted mass $-\imath /2
$.
\end{itemize} 
The global symmetry group $H$ (which is broken down to its maximal torus by the twisted masses) is $U(L)_Q\times U(L)_{\widetilde Q}\times U(1)_B\times U(1)_{\widetilde B}$. The quiver diagram is shown in Figure~\ref{fig:quiver-B}.  In this case,
the effective twisted superpotential is given by
\begin{small}
  \begin{multline}
    \label{eq:Weff-B}
    \widetilde W_{\text{eff}}^B ( \sigma ) = \frac{L}{2\pi}
    \sum_{p=1}^{N_h + N_\da} \left[ \left( \sigma_p^{(2)} +
        \frac{\imath}{2} \right) \left( \log (\sigma_p^{(2)} +
        \frac{\imath}{2}) - 1 \right) - \left( \sigma_p^{(2)} -
        \frac{\imath}{2} \right) \left( \log (-
        \sigma_p^{(2)} + \frac{\imath}{2}) - 1 \right) \right] \\
    + \frac{1}{2\pi} \sum_{i=1}^{N_\da} \sum_{p=1}^{N_h + N_\da}
    \left[ \left( \sigma_i^{(1)} - \sigma_p^{(2)} - \frac{\imath}{2}
      \right) \left( \log ( \sigma_i^{(1)} - \sigma_p^{(2)} -
        \frac{\imath}{2} ) - 1 \right) \right. \\ \left. - \left(
        \sigma_i^{(1)} - \sigma_p^{(2)} + \frac{\imath}{2} \right)
      \left( \log ( - \sigma_i^{(1)} + \sigma_p^{(2)} -
        \frac{\imath}{2} ) - 1 \right) \right] - \imath \tau_1
    \sum_{i=1}^{N_\da} \sigma^{(1)}_i - \imath \tau_2
    \sum_{p=1}^{N_h+N_\da} \sigma^{(2)}_p \, .
  \end{multline}
\end{small}
The twisted masses are compatible with a superpotential of the type
\begin{equation}
  W_B (Q^2,\overline{Q^2}, B^{12},  B^{21}) = a \sum_{k} \left[ Q^2_{\phantom{2}k} B^{21}  B^{12} \overline{Q^2}_k \right] \, .
\end{equation}

\paragraph{Case C} A quiver gauge theory with gauge groups $U(N_e)$
and $U(N_\da)$, with the following matter content:
\begin{itemize}
\item a bifundamental field $B^{12}$ in the representation
  $\overline{\mathbf{N_\da}}\otimes \mathbf{N_e}$ with twisted mass
  $\imath/2$,
\item a bifundamental field $B^{21}$ in the representation
  $\overline{\mathbf{N_e}}\otimes \mathbf{N_\da}$ with twisted mass
  $\imath/2$,
\item an adjoint field $\Phi^1$ for $U(N_\da)$ with mass $-\imath$
\item $L$ fundamental and anti--fundamentals fields
  $(Q^2_{\phantom{2}k}, \overline{Q^2}_k)$ for $U(N_e)$ with mass
  $-\imath /2 $.
\end{itemize}
The global symmetry group $H$ (which is broken down to its maximal
torus by the twisted masses) is $U(L)_Q\times U(L)_{\widetilde
  Q}\times U(1)_B\times U(1)_{\widetilde B}\times U(1)_\Phi$. The
quiver diagram is given in Figure~\ref{fig:quiver-C}. The effective
twisted superpotential reads
\begin{small}
  \begin{multline}
    \label{eq:Weff-C}
    \widetilde W_{\text{eff}}^C ( \sigma ) = \frac{L}{2\pi}
    \sum_{p=1}^{N_e} \left[ \left( \sigma_p^{(2)} + \frac{\imath}{2}
      \right) \left( \log (\sigma_p^{(2)} + \frac{\imath}{2}) - 1
      \right) - \left( \sigma_p^{(2)} - \frac{\imath}{2} \right)
      \left( \log (-
        \sigma_p^{(2)} + \frac{\imath}{2}) - 1 \right) \right] \\
    + \frac{1}{2\pi}\sum_{i=1}^{N_\da} \sum_{p=1}^{N_e} \left[ \left(
        \sigma_i^{(1)} - \sigma_p^{(2)} - \frac{\imath}{2} \right)
      \left( \log ( \sigma_i^{(1)} - \sigma_p^{(2)} - \frac{\imath}{2}
        ) - 1 \right) \right. \\ \left. - \left( \sigma_i^{(1)} -
        \sigma_p^{(2)} + \frac{\imath}{2} \right) \left( \log ( -
        \sigma_i^{(1)} + \sigma_p^{(2)} - \frac{\imath}{2} ) - 1
      \right) \right] \\ + \frac{1}{2\pi}\sum_{\substack{i,j\\i \neq j
      }}^{N_\da} \left( \sigma_i^{(1)} - \sigma_j^{(1)} + \imath
    \right) \left( \log (\sigma_i^{(1)} - \sigma_j^{(1)} + \imath ) -
      1 \right) - \imath \tau_1 \sum_{i=1}^{N_\da} \sigma^{(1)}_i -
    \imath \tau_2 \sum_{p=1}^{N_e} \sigma^{(2)}_p \, .
  \end{multline}
\end{small}
The twisted masses are compatible with a superpotential of the type
\begin{equation}
  W_C (Q^2,\overline{Q^2}, \Phi^1, B^{12},  B^{21}) = \sum_{k} \left[ a \, B^{21} \Phi^1 B^{12} + b \, Q^2_{\phantom{2}k} B^{21}  B^{12} \overline{Q^2}_k \right] \, .
\end{equation}

Even though these three theories have different gauge groups and field content, we will show with the help of the Gauge/Bethe correspondence that their supersymmetric ground states are the same.


\section{Gauge/Bethe correspondence}
\label{sec:correspondence}

The Gauge/Bethe correspondence, as detailed in~\cite{Nekrasov:2009uh,
  Nekrasov:2009ui}, relates two dimensional $\mathcal{N}=(2,2)$
supersymmetric gauge theories to quantum integrable systems. The
supersymmetric vacua of the gauge theories form a representation of
the \emph{chiral ring}, which is a distinguished class of operators
which are annihilated by one chirality of the supercharges $Q$. The
\emph{commuting Hamiltonians} of the quantum integrable system are
identified with the generators of the chiral ring. The space of states
of the quantum integrable system, \emph{i.e.} the spectrum of the
commuting Hamiltonians, is thus mapped to the supersymmetric vacua of
the gauge theory.  \changed Arguably, this correspondence holds true for all integrable
systems, in the sense that to any spin chain solvable by the Bethe
ansatz, we can associate a corresponding $\mathcal{N}=(2,2)$ gauge
theory.

\subsection{Parameters of a general spin chain}
\label{sec:parameters}

In this section, we collect the possible parameters of the quantum integrable
systems which we will need to match to the parameters of the
$\mathcal{N}=(2,2)$ gauge theories. We will be very brief; for more
detail, we refer the reader to the original
work~\cite{Nekrasov:2009uh}. We are only considering
integrable systems which correspond to two--dimensional gauge
theories, therefore we only look at spin chains without anisotropy.\footnote{By the term anisotropy we refer to the spin interactions in the Hamiltonian not being the same in the $x$, $y$ and $z$ directions, as is the case in the \textsc{xxz} and \textsc{xyz} models.}

Quantum integrable systems in $1+1$ dimensions usually correspond to
spin chain--type systems. Such a system lives on a one--dimensional
lattice of \emph{length} $L$. To each point $k$ we associate a
representation $\Lambda$ of the symmetry group $K$ and call the
corresponding Hilbert space $\mathscr{H}_k$. The dynamics is
described by the Hamiltonian
\begin{equation}
  \label{eq:spin-Hamiltonian}
  \HH = -  \sum_{k=1}^{L-1} \left[ \Pi_{k, k+1} - 1 \right] \, , 
\end{equation}
where $\Pi_{k,k+1}$ is the permutation operator between the points $k$
and $k+1$.  Moreover, one needs to specify \emph{boundary conditions}
via an operator $\mathcal{K} \in \operatorname{End}
(\mathscr{H})$. For a closed spin chain, $\mathcal{K}$ depends on
$r=\rank (K)$ \emph{twist parameters} $\set{\hat \vartheta_a}_{a=1}^r$.

Since $\HH$ commutes with the maximal
torus $T \subset K$ (summed over the chain), we can decompose the
Hilbert space of states into a direct sum
\begin{equation}
  \mathscr{H} = \bigotimes_{k=1}^L \mathscr{H}_k = \bigoplus_{a=1}^r \bigoplus_{N_a=0}^L \mathscr{H}^{(a)}_{N_a} \, .
\end{equation}
An element $\Psi \in \mathscr{H}^{(a)}_{N_a}$ is a magnon describing a
state with $N_a$ particles of \emph{species} $a$. The magnon $\Psi$ depends on the \emph{rapidities} (quasi--momenta)
$\set{\lambda^{(a)}_i}_{i = 1}^{N_a}$. There can be different
effective lengths $L_a$ for each species.  For a general spin chain,
each point $k=1, \dots, L$ can carry a different representation
$\Lambda_k = [\Lambda^{1}_k, \dots \Lambda^{r}_k]$ of the symmetry
group, and furthermore, one can turn on \emph{inhomogeneities}
$\nu^{(a)}_k$ (which can be understood as displacements) in each
position of the chain.
Special cases are the \textsc{xxx} spin chain, where each point carries
the fundamental representation of $su(2)$, and the $tJ$ model where
each point carries the fundamental representation of $sl (1|2)$.

\subsection{Algebraic Bethe ansatz}
\label{sec:algebr-bethe-ansatz}

In this section, we introduce some necessary definitions from the
theory of integrable models. In particular, we show how to construct a
system of commuting Hamiltonians starting from the Yang--Baxter
relations for a graded (supersymmetric) vector space
$\setC^{(m|n)}$. In the special case $\setC^{(1|2)}$, the construction
provides the Bethe ansatz for the $tJ$ model, which is related via the
Gauge/Bethe correspondence to the three quiver gauge theories
introduced in Sec.~\ref{sec:examples}. A pedagogical introduction to
the algebraic Bethe ansatz can be found in~\cite{Faddeev:1996iy}.

Consider a homogeneous chain of length $L$ where each position carries the
fundamental representation of the algebra $sl(m|n)$, \emph{i.e.} we have
$m$ bosonic and $n$ fermionic degrees of freedom. Associate to each
point a copy of the Hilbert space $\mathscr{H} = \setC^{(m|n)}$, where
$\setC^{(m|n)} $ is a $\setZ_2$--graded vector space $\setC^{(m|n)} =
\setC^m \oplus \setC^n$ with parity
\begin{equation}
  \abs{x} =
  \begin{cases}
    0 & \text{if $x \in \setC^m$} \\
    1 & \text{if $x \in \setC^n$.}
  \end{cases}
\end{equation}
Introduce now the matrix (linear operator on $\setC^{(m|n)} \otimes
\setC^{(m|n)}$) $R(\lambda)$, depending on the \emph{spectral
  parameter} $\lambda$. The matrix $R$ satisfies the \emph{Yang--Baxter
equation} (\textsc{ybe}) if the following identity (on $\setC^{(m|n)}
\otimes \setC^{(m|n)} \otimes \setC^{(m|n)}$) holds:
\begin{equation}
  \left( \un \otimes R(\lambda - \mu) \right) \left( R(\lambda) \otimes \un \right) \left( \un \otimes R(\mu) \right) = \left( R(\mu) \otimes \un \right)  \left( \un \otimes R(\lambda) \right)  \left( R(\lambda - \mu) \otimes \un \right) \, .
\end{equation}
The solution to the \textsc{ybe} can be written in terms of
the identity and the permutation operator
\begin{equation}
  R(\lambda) = \frac{\imath}{\lambda + \imath} \un + \frac{\lambda}{\lambda + \imath} \,\Pi \, .
\end{equation}
Moreover, the \textsc{ybe} can be rewritten in the form
\begin{equation}
  R_{12} (\lambda - \mu) \left( \Pi_{13} R_{13} (\lambda) \otimes \Pi_{23} R_{23} (\mu) \right) = \left( \Pi_{13} R_{13} (\mu) \otimes \Pi_{23} R_{23}(\lambda) \right) R_{12}(\lambda-\mu) \, , 
\end{equation}
where the indices $1,2,3$ indicate on which of the three
$\setC^{(m|n)}$ spaces the operator is acting. In this form, we have
singled out the third space. Now we can choose to consider it
differently from the other two and interpret it as the Hilbert space
$\mathscr{H}_k$ living on a point of the chain while the others take
the role of auxiliary spaces. In turn, we can require a \textsc{ybe}
to be satisfied at each point. It is convenient to introduce the
\emph{Lax operator} $L_k$ at the point $k$ as follows:
\begin{equation}
  L_k (\lambda) = \Pi \,R(\lambda) = \frac{\imath}{\lambda + \imath} \Pi + \frac{\lambda}{\lambda + \imath} \un \, .  
\end{equation}
Since $\mathscr{H}$ is an inner quantum space, $L_k(\lambda)$ can now
be seen as an $(m+n) \times (m+n)$ matrix, whose entries are quantum
operators. The \textsc{ybe} at point $k$ is written as
\begin{equation}
  R(\lambda - \mu) \left( L_k (\lambda) \otimes L_k (\mu) \right) = \left( L_k (\mu) \otimes L_k(\lambda) \right)  R(\lambda - \mu ) \, .
\end{equation}
In physical terms, we can interpret the Lax operator as a
connection along the chain, in the sense that $L_k$ defines the
transport between the points $k$ and $k+1$, via the \emph{Lax
  equation}:
\begin{equation}
  \Psi_{k+1} = L_k \Psi_k \, ,  
\end{equation}
where the vector $\Psi$ has $m+n$ entries in $\mathscr{H}$.  It is
thus natural to define the monodromy matrix $T(\lambda)$ as the
ordered product of $L_k(\lambda)$ along the chain:
\begin{equation}
  T(\lambda) = L_L (\lambda) L_{L-1} (\lambda) \cdots L_1(\lambda) \, . 
\end{equation}
One can show by induction that the monodromy matrix
satisfies the same \textsc{ybe}:
\begin{equation}
  R(\lambda - \mu) \left( T (\lambda) \otimes T (\mu) \right) = \left( T (\mu) \otimes T(\lambda) \right)  R(\lambda - \mu ) \, .
\end{equation}
Taking the (super) trace of $T(\lambda)$ in the auxiliary space gives
the \emph{transfer matrix} $t(\lambda)$:
\begin{equation}
  t(\lambda) = \Tr [T(\lambda) ] \, ,  
\end{equation}
the \textsc{ybe} implies that transfer matrices at
different values of the spectral parameter commute:
\begin{equation}
  [ t(\lambda), t(\mu )] = 0 \, .  
\end{equation}
This is the fundamental property that turns $t(\lambda)$ into the
generating object for the $L-1$ integrals of motion that make the system
integrable. It is customary to define the $L-1$ Hamiltonians
$\HH_{(l)}$ as the coefficients of the development
\begin{equation}
  \log [ t(\lambda) t(0)^{-1}] = \sum_{l=1}^L \imath \frac{\lambda^l}{l!} \HH_{(l)} \, .  
\end{equation}
In particular, the logarithmic derivative of $t(\lambda)$ in $\lambda
= 0$ is the Hamiltonian for the spin chain that we introduced in
Eq.~(\ref{eq:spin-Hamiltonian}):
\begin{equation}
  \HH = \HH_{(1)} = - \imath \left. \frac{\di}{\di\lambda} \log t(\lambda) \right|_{\lambda=0} =  -  \sum_{k=1}^{L-1} \left[ \Pi_{k, k+1} - 1 \right] \, .
\end{equation}
Eigenvectors for the transfer matrix are at the same time eigenvectors
for all the commuting Hamiltonians. In order to construct them,
consider the simplest case $n=2,\ m=0$, corresponding to the
\textsc{xxx} spin chain. The monodromy can be written as an operator--valued $2\times2$ matrix:
\begin{equation}
  T (\lambda) =
  \begin{pmatrix}
    A(\lambda) & B(\lambda) \\
    C(\lambda) & D(\lambda) 
  \end{pmatrix} \, .
\end{equation}
Introduce a reference state $\Omega$ such that
\begin{align}
  \Omega &= \bigotimes_{k=1}^L \omega_k \, ,& L_k \omega_k &=
  \begin{pmatrix}
    \alpha(\lambda) & * \\ 0 & \delta(\lambda)
  \end{pmatrix} \omega_k & \Rightarrow &&
T(\lambda) \Omega &=
  \begin{pmatrix}
    \alpha (\lambda)^L & * \\
    0 & \delta(\lambda)^L
  \end{pmatrix} \Omega \, ,
\end{align}
where $\alpha (\lambda) = \lambda + \imath /2 $ and $\delta(\lambda) =
\lambda - \imath / 2 $, then by construction $\Omega$ is an eigenvector for $t(\lambda)$:
\begin{equation}
  t(\lambda)\, \Omega = \left( \alpha (\lambda)^L + \delta(\lambda)^L \right) \Omega \, .
\end{equation}
The \emph{algebraic Bethe ansatz} consists in looking for eigenvectors
of $t(\lambda)$ of the form
\begin{equation}
  \Phi (\{ \lambda\} ) = B(\lambda_1) B(\lambda_2) \cdots B(\lambda_N)\, \Omega \, .  
\end{equation}
Imposing the \textsc{ybe} for the monodromy matrix, one finds that
$\Phi(\{ \lambda\})$ is an eigenvector if and only if the parameters
$\{ \lambda\}$ satisfy the Bethe equations:
\begin{equation}
  \left( \frac{ \lambda_i + \frac{\imath}{2} }{ \lambda_i - \frac{\imath}{2} }\right)^L = \prod_{\substack{j=1\\j\neq i}}^N\, \frac{ \lambda_i - \lambda_j + \imath}{\lambda_i - \lambda_j - \imath} \, , \hspace{2em} i= 1,2, \dots, N \, .
\end{equation}

The case of a  higher rank symmetry group is analyzed with a recursive procedure known as
the \emph{nested Bethe ansatz}
(see~\cite{KulishReshetikhin,Kazakov:2007fy}). For $n+m > 2$, applying
the construction above produces, together with a set of Bethe
equations for variables $\{\lambda^{(1)}\}$, a new Lax operator with rank
$n+m -1 $. The construction can now be repeated, and at each step one
obtains a system of equations for a new set of variables
$\{\lambda^{(a)}\}$ and a new Lax operator of rank $n+m-a$. After $n + m -
2 $ steps, one arrives at the final set of equations. The overall
result can be put into a very elegant form in which the Bethe
equations only depend on the root space decomposition of the symmetry
group~\cite{Reshetikhin:1987xx}. Explicitly,
\begin{equation}
  \label{eq:general-NBAE}
  \left( \frac{ \lambda_i^{(a)} + \frac{\imath}{2} \Lambda^{a} }{ \lambda_i^{(a)} - \frac{\imath}{2} \Lambda^a }\right)^L = \prod_{\substack{(b,j)=(1,1)\\(b,j)\neq (a,i)}}^{(r,N_b)} \frac{ \lambda^{(a)}_i - \lambda^{(b)}_j + \frac{\imath}{2} C^{ab}}{\lambda^{(a)}_i - \lambda^{(b)}_j - \frac{\imath}{2} C^{ab}} \, , \hspace{2em} a=1,2, \dots, r \, , \hspace{1em} i= 1,2, \dots, N_a \, ,
\end{equation}
where $r = \rank ( sl(m|n) ) = n + m - 1 $, $C^{ab}$ is Cartan matrix,
and $[\Lambda^1, \dots, \Lambda^r]$ is the highest weight of the
representation\footnote{In this notation, the spin $\frac{1}{2}$
  representation of $su(2)$ of the ``standard'' \textsc{xxx} model has
  weight $[\Lambda^1] =[1] $.}.
Two remarks are of importance:
\begin{enumerate}
\item Even though we started by considering the fundamental representation
  for $sl(m|n)$, the Bethe ansatz equations~\eqref{eq:general-NBAE}
  are more general and valid for an arbitrary representation
  $\Lambda$ of the symmetry group $K$.
\item The result depends on the choice of the Cartan matrix $C^{ab}$ of the
  symmetry group. While this is unique (up to conjugation) for $sl(n)$,
  this is not the case for $sl(m|n)$, where there are $\binom{m+n}{m}$
  conjugacy classes of Borel subalgebras. This means that for a given
  spin chain (and for a given ring of commuting Hamiltonians), there
  are $\binom{m+n}{m}$ sets of Bethe ansatz equations which are by
  construction equivalent.
\end{enumerate}
We would like to stress that for a given spin chain with supergroup
symmetry, there is a unique ring of commuting Hamiltonians, but the
choice of Borel subalgebra can lead to different--looking Bethe
equations. This is the property that we will use in the following to
prove that different quiver gauge theories (one for each choice of
Bethe equations) have the same ground states (which are ultimately
identified by the ring of Hamiltonians).

\bigskip

A very non--trivial statement is that the Bethe equations
describe the critical points of a potential, which was first
introduced in~\cite{Yang:1968rm} by Yang and Yang. The
\emph{Yang--Yang function} corresponding to the nested Bethe ansatz
equations in Eq.~\eqref{eq:general-NBAE} reads:
\begin{equation}
  \label{eq:Yang-Yang-NBAE}
  Y ( \lambda) = \frac{L}{2\pi} \sum_{a=1}^{r} \Lambda^a \sum_{i=1}^{N_a} \hat x ( \tfrac{2\lambda^{(a)}_i}{\Lambda^a} ) - \frac{1}{4\pi} \sum_{a,b=1}^r C^{ab} \sum_{(i,j)=(1,1)}^{(N_a, N_b)} \hat x ( \tfrac{2\lambda^{(a)}_i - 2\lambda^{(b)}_j}{C^{ab}})  + \sum_{a=1}^r \sum_{i=1}^{N_a} n^{(a)}_i \lambda^{(a)}_i,
\end{equation}
where the $n_i^{(a)}$ are integers and $\hat x$ is the function
\begin{equation}
  \hat x (\lambda) = \lambda \arctan (\lambda^{-1}) + \frac{1}{2} \log(1+\lambda^2) \, ,
\end{equation}
which satisfies
\begin{equation}
  e^{2 \imath x^\prime ( a \lambda)} = \frac{\lambda+ \imath/a}{\lambda - \imath/a} \, .  
\end{equation}
It follows that the system of Bethe equations in
Eq.~\eqref{eq:general-NBAE} can be written as
\begin{equation}
  \label{eq:Bethe-as-exponential}
  e^{ 2 \pi \imath \varpi^{(a)}_i ( \lambda )} = 1  \, , \hspace{2em} a=1,2, \dots, r \, , \hspace{1em} i= 1,2, \dots, N_a \, ,
\end{equation}
where $\varpi_i^{(a)}(\lambda) $ are the components of the
closed one--form $  \varpi (\lambda) = \di Y (\lambda) $,
\begin{equation}
  \varpi(\lambda) = \sum_{a=1}^r \sum_{i=1}^{N_a} \varpi(\lambda)_i^{(a)} \di \lambda_i^{(a)} = \sum_{a=1}^r \sum_{i=1}^{N_a} \frac{\partial Y(\lambda)}{\partial \lambda_i^{(a)}}  \di \lambda_i^{(a)} = \di Y(\lambda)\, .
\end{equation}

\subsection{The Dictionary}
\label{sec:dictionary}

The main statement of~\cite{Nekrasov:2009uh,Nekrasov:2009ui} is that
the effective twisted superpotential $\widetilde
W_{\text{eff}}(\sigma)$ can be identified with the Yang--Yang counting
function $Y(\lambda)$, once the parameters of both theories are
properly matched. In this section, finally, we give the precise
dictionary between the quantities of the ${\mathcal N}=(2,2)$ gauge
theory and the integrable systems we have introduced.

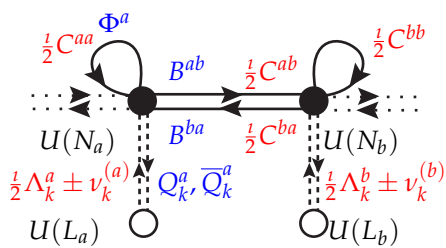
\begin{figure}
  \centering
  \begin{picture}(140,100)(,)
      \begin{small}
        \put(8,0){$U(L_a)$} \put(120,0){$U(L_b)$} 
        \put(12,34){$U(N_a)$} \put(118,34){$U(N_b)$}
        \textcolor{red}{\put(0,18){$\frac{\imath}{2}\Lambda^a_k \pm \nu^{(a)}_k$} \put(118,18){$\frac{\imath}{2}\Lambda^b_k \pm \nu^{(b)}_k$} 
          \put(12,70){$\frac{\imath}{2}C^{aa}$} \put(88,60){$\frac{\imath}{2}C^{ab}$} \put(88,36){$\frac{\imath}{2}C^{ba}$} \put(136,72){$\frac{\imath}{2}C^{bb}$}
        \textcolor{blue}{\put(56,18){$Q^a_k, \overline{Q}^a_k$}
        \put(34,78){$\Phi^{a}$} \put(60,60){$B^{ab}$} \put(60,36){$ B^{ba}$} 
      }
      }
      \end{small}
      \includegraphics[scale=1]{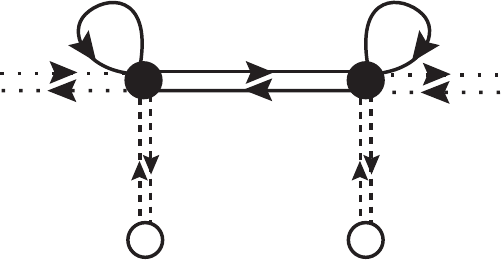}
    \end{picture}
    \caption{Example quiver diagram for the Gauge/Bethe
      correspondence. Gauge groups are labeled in black, matter fields in blue, the corresponding twisted
      masses in red. }
  \label{fig:dictionary}
\end{figure}

The first observation is that the equation~\eqref{eq:vacua} for the
vacua of the gauge theory and the Bethe ansatz
equation~\eqref{eq:Bethe-as-exponential} for the rapidities have the
same form.  Most properties of the gauge theory are determined by the
symmetry group $K$ of the integrable system. The sector with particle
numbers $\{N_a\}_{a=1}^r$ for each species leads to a product gauge group of the form
$\prod_{a=1}^r U(N_a)$. This results in a quiver gauge theory with $r$
nodes, where the node $a$ carries the gauge group $U(N_a)$. Each
effective length $L_a$ gives rise to $L_a$ fundamentals and $L_a$
anti--fundamental fields being attached to node $a$. The twisted
masses of the bifundamental and adjoint fields can be read off from
the Cartan matrix of $K$. In the quiver diagram, we only draw those lines between nodes $a$,
$b$ which correspond to a non--zero entry $C^{ab}$ (\ie
to non--zero twisted mass). We are thus lead to a quiver diagram of
the type shown in Figure~\ref{fig:dictionary}.
The twisted masses of the $k$--th fundamental and anti--fundamental field at
node $a$ are given by the weight of the representation of the symmetry group $K$
that the position $k$ in the chain is carrying, plus the possible
inhomogeneity at position $k$.  The boundary conditions for closed spin
chains, which are encoded in the $\hat \vartheta^a$, enter the FI terms of the gauge
theory.\footnote{Periodic spin chains give rise to $U(N)$ gauge groups,
  while open chains result in $SO(N)$ or $Sp(N)$ gauge groups,
  depending on the boundary condition. The boundary conditions for open spin chains are not described by $\hat \vartheta^a$--parameters, which corresponds to the fact that the $SO(N)$ and $Sp(N)$ groups do not have a central $U(1)$--factor and thus have no FI--terms.}

The Coulomb branch only depends on the effective twisted
superpotential and is not affected by the presence of an
$F$--term. Nevertheless, in general the superpotential will break
(part of) the global symmetries which results in
constraints on the possible values of the twisted masses. These constraints are
to be compared with those that come from
the theory of representations of the symmetry group $K$ on the integrable model side (\emph{e.g.} the Cartan matrix containing only integer entries, or the
allowed values for the highest weights).

All the relevant parameters and their matching are collected in Table~\ref{tab:dictionary}.

\renewcommand{\arraystretch}{1.3}

\newcolumntype{S}{>{\raggedright\arraybackslash}m{11.5em}}
\newcolumntype{T}{>{\arraybackslash}m{14em}}

\begin{table}[t]
  \centering
  \begin{tabular}{SccT}
    \toprule
    \multicolumn{2}{c}{gauge theory} & \multicolumn{2}{c}{integrable model} \\ \midrule
    number of nodes in the quiver & $r$ & $r$ & rank of the symmetry group \\
    gauge group at $a$--th node &$U(N_a)$ & $N_a$ & number of particles of species $a$ \\
    effective twisted super\-potential & $\widetilde W_{\text{eff}}(\sigma)$ & $Y(\lambda)$ & Yang--Yang function \\
    equation for the vacua & $e^{2 \pi \di \widetilde W_{\text{eff}}} = 1 $& $e^{2 \pi \imath \di Y} = 1 $ & Bethe ansatz equation \\ 
    flavor group at node $a$& $U(L_a)$ & $L_a$ & effective length for the species $a$ \\
    lowest component of the twisted chiral superfield &$\sigma^{(a)}_i$ & $\lambda^{(a)}_i$ & rapidity \\
    twisted mass of the fundamental field & ${{}\widetilde{m}^{\fund}}_{k}^{(a)}$ & $\frac{\imath}{2} \Lambda^{a}_k + \nu^{(a)}_k$ & highest weight of the representation  and inhomogeneity \\
    twisted mass of the anti--fundamental field & ${{}\widetilde{m}^{\bar \fund}}^{(a)}_{k}$ & $\frac{\imath}{2} \Lambda^{a}_k - \nu^{(a)}_k$ & highest weight of the representation  and inhomogeneity \\
    twisted mass of the adjoint field & ${{}\widetilde{m}^{\adj}}^{(a)}$ & $\frac{\imath}{2} C^{aa}$ & diagonal element of the Cartan matrix \\
    twisted mass of the bifundamental field & ${{}\widetilde{m}^{\bif}}^{(ab)}$ & $\frac{\imath}{2} C^{ab} $& non--diagonal element of the Cartan matrix \\
    FI--term for $U(1)$--factor of gauge group $U(N_a)$ & $\tau_a$&$\hat\vartheta^a$&boundary twist parameter for particle species $a$\\

    \bottomrule
  \end{tabular}
  \caption{Dictionary in the Gauge/Bethe correspondence.}
  \label{tab:dictionary}
\end{table}

\subsection{Example: XXX spin chain}
\label{sec:duality-xxx-spin}
\changed

The \textsc{xxx} spin chain is one of the best studied integrable
models. It describes a system of electrons on a lattice with spin
exchange interactions. Each site can be either occupied by a spin up
($\ua$) or down ($\da$). The Hilbert space at each point is
\begin{equation}
  \mathscr{H}_k = \setC^2 \, , 
\end{equation}
which corresponds to the fundamental representation of $sl(2)$. Using
the results of the previous section and choosing as a reference state
\begin{equation}
  \Omega = \bigotimes_{i=1}^L e_{\da} \, , 
\end{equation}
one finds that the rapidities $\lambda$ satisfy the Bethe Ansatz
equations
\begin{equation}
  \left( \frac{\lambda_i + \frac{\imath}{2}}{\lambda_i -
      \frac{\imath}{2}} \right)^L e^{\imath \hat \theta} =
  \prod_{\substack{j=1\\ j \neq i}}^{N}\frac{\lambda_i-\lambda_j +
    i}{\lambda_i-\lambda_j - i}, \quad i=1,\dots,N \, ,
\end{equation}
where $L$ is the length of the chain, $N$ is the number of magnons and
$\hat \theta$ is the boundary twist parameter.

It is a known fact (for a modern discussion see~\cite{Bazhanov:2010ts})
that there is a completely equivalent set of equations obtained by
choosing the opposite reference state
\begin{equation}
  \Omega = \bigotimes_{i=1}^L e_{\ua}
\end{equation}
and considering $L-N$ dual magnons:
\begin{equation}
  \left( \frac{\lambda_i + \frac{\imath}{2}}{\lambda_i -
      \frac{\imath}{2}} \right)^L e^{-\imath \hat \theta} =
  \prod_{\substack{j=1\\ j \neq i}}^{L-N}\frac{\lambda_i-\lambda_j +
    i}{\lambda_i-\lambda_j - i}, \quad i=1,\dots,N \, .  
\end{equation}

Comparing the Bethe Ansatz equations and their corresponding
Yang--Yang functions to the effective twisted superpotentials in
Eq.~\eqref{eq:XXX-1} and Eq.~\eqref{eq:XXX-2} we find that the two
gauge systems described in Sec.~\ref{sec:example:-} admit the same $2^L$
supersymmetric ground states.

\subsection{Example: $tJ$ model}
\label{sec:example:-tj-model}

The $tJ$ model~\cite{PhysRevB.37.3759} describes a system of electrons
on a lattice with a Hamiltonian that describes nearest--neighbor
hopping (with coupling $t$) and spin interactions (with coupling
$J$). Consider a lattice of length $L$ with periodic boundary conditions. Each site can be either free
($\circ$) or occupied by a spin up ($\ua$) or down ($\da$)
electron. Excluding double occupancy, the Hilbert space at each point
$k$ is:
\begin{equation}
  \mathscr{H}_k = \setC^{(1|2)} \, ,
\end{equation}
which corresponds to the fundamental representation of $sl(1|2)$.  It
is convenient to introduce anticommuting creation--annihilation pairs
$c^\dagger_{k, s}, c_{k,s}$, $s = \set{\ua, \da} $ at
each site, acting as
\begin{align}
  \ket{s}_k &= c^\dagger_{k,s} \ket{\circ}_k \, , & \text{for $s = \set{\ua, \da}$,}  
\end{align}
where $\ket{\circ}_k$ is the vacuum, annihilated by $c_{k,
  s}$. Let $n_{k, s} = c^\dagger_{k,s} c_{k,s}$ be
the number of $s$ electrons at position $k$ and $n_k = n_{k,\ua}
+ n_{k, \da}$. We can further introduce $sl(2)$ spin operators at each
site:
\begin{align}
  S^-_k = c^\dagger_{k,\ua} c_{k,\da} \, , && S_k^+ = c^{\dagger}_{k,\da} c_{k,\ua} \, , && S_k^z = \frac{1}{2} \left( n_{k,\ua} - n_{k,\da} \right) \, .
\end{align}
With these ingredients, we can write down the Hamiltonian
\begin{equation}
  \HH = \sum_{k=1}^{L-1} \left[ -t\, \mathcal{P} \sum_{s={\ua, \da}} \left( c^\dagger_{k, s} c_{k+1,s}  + \text{h.c.} \right) \mathcal{P} + J \left( \mathbf{S}_k \cdot \mathbf{S}_{k+1} - \frac{1}{4} n_k n_{k+1} +\,2\, n_k - \frac{1}{2} \right)\right] \, ,
\end{equation}
where $\mathcal{P}$ projects out double occupation. It is
convenient to introduce the number of holes $N_h$, of spins up
$N_\ua$, of spins down $N_\da$, and electrons $N_e$:
\begin{align}
  N_h &= \sum_{k=1}^L ( 1 - n_k) \, ,& N_\ua &= \sum_{k=1}^L n_{k,\ua} \, ,& N_{\da} &= \sum_{k=1}^L n_{k,\da} \, ,& N_e &= N_\ua + N_\da \, .   
\end{align}
Single occupancy implies
\begin{equation}
  L = N_h + N_\ua + N_\da \, .  
\end{equation}

The Hamiltonian is remarkable for being supersymmetric for the choice
$J = 2t =2 $, in the sense that it is invariant under the action of
the superalgebra $sl(1|2)$ (see Appendix~\ref{sec:superalgebra} for an
explicit realization of the algebra in terms of creation/annihilation
operators and for some basic properties). In the supersymmetric
case, $\HH$ can also be conveniently expressed as
\begin{equation}
  \HH = - \sum_{k=1}^{L-1} \left[ \Pi_{k,k+1} - 1  \right] \, ,
\end{equation}
where $\Pi_{k,k+1}$ interchanges the configurations at sites $k $ and
$k+1$, with an extra minus sign if they are both fermionic:
\begin{align}
  \Pi_{k,k+1} \ket{\circ}_k \otimes \ket{\circ}_{k+1} &= \ket{\circ}_k \otimes \ket{\circ}_{k+1} \, , \\
  \Pi_{k,k+1} \ket{\circ}_k \otimes \ket{s}_{k+1} &= \ket{s}_k \otimes \ket{\circ}_{k+1} \, , &s&= \set{\ua,\da} \,,  \\
  \Pi_{k,k+1} \ket{s_1}_k \otimes \ket{s_2}_{k+1} &= - \ket{s_2}_k \otimes \ket{s_1}_{k+1}  \, , & s_1, s_2 &= \set{\ua,\da} \, .
\end{align}
This is precisely the same structure that we introduced in the
previous section.

\bigskip 

Having recognized the $tJ$ model Hamiltonian as an example of a spin
chain solvable by the algebraic Bethe ansatz, we can use the results
of Sec.~\ref{sec:algebr-bethe-ansatz}.  A fundamental remark is in
order. This spin chain is $sl(1|2)$--invariant, and this supergroup
admits different inequivalent choices of the Cartan matrix (as shown
in Appendix~\ref{sec:superalgebra}).  This means that the same
physical system of electrons and holes is described by three different
Bethe ansatz equations, as explained in~\cite{Essler:1992he}. By
construction, the three choices must be equivalent, as was shown
explicitly in~\cite{Essler:1992he,PhysRevB.46.14624} (see also
Appendix~\ref{sec:equivalence}).  We will examine all three of them
here.

\paragraph{Case A}

The first case corresponds the Kac--Dynkin\footnote{See Appendix~\ref{sec:superalgebra}.} diagram $\diagA{0}{1}$. This
leads to
\begin{align}
  C^{ab} &=
  \begin{pmatrix}
    0 & -1 \\ -1 & 2
  \end{pmatrix} \, , & \Lambda &= [ 0 \ 1] \, , & N_1 &= N_h \, , & N_2 &= N_h + N_\da \, .
\end{align}
The nested Bethe equations are given by
\begin{subequations}
\label{eq:Bethe-A}
  \begin{align}
    \left(\frac{\lambda^{(2)}_p+\frac{i}{2}}{\lambda^{(2)}_p-\frac{i}{2}}\right)^L
    &= \prod_{\substack{q=1\\ q\neq p}}^{N_h+N_{\da}}\frac{
      \lambda^{(2)}_p - \lambda^{(2)}_q + {i}}{\lambda^{(2)}_p -
      \lambda^{(2)}_q - {i}}\,\prod_{i=1}^{N_h} \frac{\lambda^{(2)}_p -
      \lambda_i^{(1)} - \frac{i}{2}}{\lambda^{(2)}_p - 
      \lambda_i^{(1)}+\frac{i}{2}},
    \quad p=1,\dots,N_h+N_{\da}, \\
    1 &=
    \prod_{p=1}^{N_h+N_{\da}}\frac{\lambda^{(2)}_p-\lambda_i^{(1)}-\frac{i}{2}}{\lambda^{(2)}_p
      - \lambda_i^{(1)}+\frac{i}{2}}, \quad i=1,\dots,N_{h}.
  \end{align}
\end{subequations}
The Yang--Yang function~\eqref{eq:Yang-Yang-NBAE} reads:
\begin{multline}
\label{eq:Yang-Yang-A}
Y_A (\lambda) = \frac{L}{2\pi} \sum_{p=1}^{N_h +
  N_\da} \hat x ( 2 \lambda^{(2)}_p ) - \frac{1}{2\pi}\sum_{\substack{p,q=1 \\ p \neq q }}^{N_h + N_\da}
  \hat x ( \lambda^{(2)}_p - \lambda^{(2)}_q ) 
  + \frac{1}{2\pi}\sum_{p=1}^{N_h + N_\da} \sum_{i=1}^{N_h} \hat x ( 2 \lambda^{(2)}_p - 2 \lambda_i^{(1)} ) \\  + 
  \sum_{i=1}^{N_h} n_i^{(1)} \lambda_i^{(1)} +
  \sum_{p=1}^{N_h + N_\da}  n^{(2)}_p \lambda^{(2)}_p \, .
\end{multline}

\paragraph{Case B}

The second case corresponds to the Kac--Dynkin diagram $\diagB{0}{1}$.  This
leads to
 \begin{align}
   C^{ab} &=
   \begin{pmatrix}
     0 & -1 \\ -1 & 0
   \end{pmatrix} \, , & \Lambda &= [ 0 \ 1] \, , & N_1 &= N_\da \, , & N_2 &= N_h + N_\da \, .
 \end{align}
The nested Bethe equations are given by
\begin{subequations}
\label{eq:Bethe-B}
  \begin{align}
    \left(\frac{\lambda^{(2)}_p + \frac{i}{2}}{\lambda^{(2)}_p - \frac{i}{2}}\right)^L&= \prod_{i=1}^{N_{\da}}\frac{\lambda^{(1)}_i - \lambda^{(2)}_p - \frac{i}{2}}{\lambda^{(1)}_i - \lambda^{(2)}_p + \frac{i}{2}} \, , \quad p = 1,\dots, N_h+N_\da \\
    1 &= \prod_{p=1}^{N_h+N_{\da}}\frac{\lambda_i^{(1)} -
      \lambda^{(2)}_p - \frac{i}{2}}{\lambda_i^{(1)}-\lambda^{(2)}_p +
      \frac{i}{2}}\, , \quad i=1,\dots,N_{\da} \, .
  \end{align}
\end{subequations}
The Yang--Yang function~\eqref{eq:Yang-Yang-NBAE} reads:
\begin{equation}
  \label{eq:Yang-Yang-B}
  Y_B (\lambda) = \frac{L}{2 \pi} \sum_{p=1}^{N_h + N_\da}
  \hat x ( 2 \lambda^{(2)}_p ) + \frac{1}{2\pi}\sum_{p=1}^{N_h + N_\da}
  \sum_{i=1}^{N_\da} \hat x ( 2 \lambda_i^{(1)} - 2\lambda^{(2)}_p ) 
  +  \sum_{i=1}^{N_\da} n_i^{(1)} \lambda_i^{(1)}  +  \sum_{p=1}^{N_h + N_\da} n^{(2)}_p \lambda^{(2)}_p 
  \, .
\end{equation}

\paragraph{Case C}

The third case corresponds to Kac--Dynkin diagram $\diagC{0}{1}$.  This
leads to
\begin{align}
  C^{ab} &=
  \begin{pmatrix}
    2 & -1 \\ -1 & 0
  \end{pmatrix} \, , & \Lambda &= [ 0 \ 1] \, , & N_1 &= N_\da \, , & N_2 &= N_\ua + N_\da \, .
\end{align}
The nested Bethe equations are given by
\begin{subequations}
\label{eq:Bethe-C}
  \begin{align}
    \left(\frac{\lambda^{(2)}_p - \frac{i}{2}} {\lambda^{(2)}_p +
        \frac{i}{2}} \right)^L &= \prod_{i=1}^{N_{\da}}
    \frac{\lambda^{(2)}_p - \lambda_i^{(1)} - \frac{i}{2}}
    {\lambda^{(2)}_p - \lambda_i^{(1)}+\frac{i}{2}},
    \quad p = 1,\dots,N_e,\\
    \prod_{p=1}^{N_{e}}\frac{\lambda^{(2)}_p-\lambda_i^{(1)}-\frac{i}{2}}{\lambda^{(2)}_p-\lambda_i^{(1)}
      + \frac{i}{2}} &= \prod_{\substack{j=1\\ j\neq
        i}}^{N_{\da}}\frac{\lambda^{(1)}_j-\lambda_i^{(1)}-
      i}{\lambda^{(1)}_j-\lambda_i^{(1)} + i}, \quad
    i=1,\dots,N_{\da}.
  \end{align}
\end{subequations}
The resulting Yang--Yang function~\eqref{eq:Yang-Yang-NBAE} is given
by
\begin{multline}
  \label{eq:Yang-Yang-C}
  Y_C(\lambda) = \frac{L}{2\pi} \sum_{p=1}^{N_e} \hat x (2
  \lambda^{(2)}_p) - \frac{1}{2\pi}\sum_{p=1}^{N_e}
  \sum_{i=1}^{N^{\da}} \hat x (2 \lambda^{(2)}_p - 2 \lambda_i^{(1)})
  + \frac{1 }{2\pi}\sum_{\substack{i,j=1\\i\neq
      j}}^{N_{\da}} \hat x (\lambda_i^{(1)} - \lambda_j ^{(1)}) \\
  + \sum_{i=1}^{N_{\da}}n_i^{(1)}\lambda_i^{(1)} + \sum_{p=1}^{N_e}
  n^{(2)}_p \lambda^{(2)}_p \, .
\end{multline}

The three systems of equations admit a number of solutions $Z(N_h,
N_\da, N_\ua)$ that can be written explicitly
as follows~\cite{PhysRevB.46.9234}:
\begin{equation}
  Z( N_h, N_\da, N_\ua ) = \sum_{q=0}^{N_h+N_\da} \frac{N_\ua - N_\da
    + 1}{N_h + N_\ua + 1 } \binom{q-1}{N_h} \binom{N_h + N_\ua + 1}{q}
  \binom{N_h - N_\da - 1}{q-1} \, . 
\end{equation}

\subsection{Gauge/Bethe correspondence for the $tJ$ model}
\label{sec:corrtJ}

After having collected the relevant quantities both for our quiver gauge theories and the $tJ$ model, we are ready to identify the effective twisted superpotentials given in Sec. \ref{sec:examples} with the Yang--Yang functions derived in the previous section. Observing that
\begin{equation}
  \frac{2 \imath}{a} \hat x (a \lambda) = \left( \lambda + \frac{\imath}{a} \right) \left( \log ( \lambda + \frac{\imath}{a} )- 1  \right) - \left( \lambda - \frac{\imath}{a} \right) \left( \log ( \lambda - \frac{\imath}{a} )- 1  \right) + \text{const.} \, , 
\end{equation}
we are now in a position to identify the gauge theories whose effective
twisted superpotentials reproduce the Yang--Yang functions above:
\begin{itemize}
\item The Yang--Yang function in Eq.~\eqref{eq:Yang-Yang-A}
  corresponds to a quiver gauge theory with the effective twisted
  superpotential given in Eq.~\eqref{eq:Weff-A} with $\vartheta$--angles
  $\vartheta_1= \left( N_h + N_\da \right) \pi $ and $\vartheta_2 =
  \left( N_h + N_\ua + 1 \right) \pi$.
\item The Yang--Yang function in Eq.~\eqref{eq:Yang-Yang-B}
  corresponds to a quiver gauge theory with the effective twisted
  superpotential given in Eq.~\eqref{eq:Weff-B} with $\vartheta_1 = \left(
    N_h + N_\da\right)\pi$ and $\vartheta_2 = \left( N_h +
    N_\ua\right)\pi $.
\item The Yang--Yang function in Eq.~\eqref{eq:Yang-Yang-C}
  corresponds to a quiver gauge theory the with effective twisted
  superpotential given in Eq.~\eqref{eq:Weff-C} with $\vartheta_1 = \left(
    N_\ua + 1 \right) \pi $ and $\vartheta_2 = \left( N_h + N_\ua
  \right) \pi $.
\end{itemize}
In Table~\ref{tab:Kac-Dynkin-quivers}, the Kac--Dynkin
diagrams and the quiver diagrams for the corresponding gauge theories are shown.

\begin{table}
  \centering
  \begin{tabular}{lll}
    \toprule
    \hspace{1em}Case A & \hspace{1em}Case B & \hspace{2em}Case C \\
    \midrule
    \hspace{.5em}$\bigdiagA{0}{1}$ & \hspace{.5em}$\bigdiagB{0}{1}$ & \hspace{1.5em}$\bigdiagC{0}{1}$ \\ [1.5em]
    \begin{picture}(130,45)(,)
      \begin{small}
        \put(0,22){$U(N_h)$} \put(59,28){$U(N_h + N_\da) $} \put(61,1){$U(L)$}
        \textcolor{red}{\put(72,42){$\imath$} \put(23,42){$\imath/2$} \put(59,17){$-\imath/2$}}
      \end{small}
      \includegraphics[scale=.7]{quiverA-crop}
    \end{picture} &
    \begin{picture}(130,45)(,)
      \begin{small}
        \put(0,22){$U(N_\da)$} \put(61,33){$U(N_h + N_\da) $} \put(61,1){$U(L)$}
         \textcolor{red}{\put(23,42){$\imath/2$} \put(59,17){$-\imath/2$}}
      \end{small}
      \includegraphics[scale=.7]{quiverB-crop}
    \end{picture} &
    \begin{picture}(110,45)(,)
      \begin{small}
        \put(5,22){$U(N_\da)$} \put(72,33){$U(N_e)$} \put(74,1){$U(L)$}
        \textcolor{red}{\put(-8,42){$-\imath$} \put(34,42){$\imath/2$} \put(71,17){$-\imath/2$}}
      \end{small}
      \includegraphics[scale=.7]{quiverC-crop}
    \end{picture}\\ 
    \bottomrule
  \end{tabular}
  \caption{    Comparing quiver diagrams for the three supersymmetric theories and Dynkin--Kac diagrams for the fundamental representation.  For each
    node in the Dynkin diagram, there is a gauge group. For each white node, there is an adjoint field. A
    flavor group is attached to the nodes with non--zero label.}
  \label{tab:Kac-Dynkin-quivers}
\end{table}

We would like to stress once more the logic behind our
construction. The $tJ$ model admits three sets of Bethe ansatz
equations corresponding to the same ring of commuting Hamiltonians. To each of
these, we associate a quiver gauge theory, according to the dictionary
in Table~\ref{tab:dictionary}. Since the commuting
Hamiltonians are the same, also the three gauge theories have the same
chiral ring and, equivalently, the same supersymmetric ground states.

\bigskip

Having considered a supergroup symmetry, we are in the position to
slightly extend the dictionary in Section~\ref{sec:dictionary}. The
quiver diagrams for the supersymmetric gauge theories are to be
compared to the Kac--Dynkin diagrams of the superalgebra. For each
node in the Dynkin diagram, there is a gauge group. Furthermore, each
white node carries an adjoint field. A flavor group is attached to the
nodes with non--zero label.

We would like to end this section with an observation concerning the
constraints on the mass parameters coming from the two sides of the
correspondence.  Consider for simplicity the case of the
\emph{distinguished Borel subalgebra} of $sl(m|n)$, whose Dynkin
diagram has $m-1 $ white nodes, followed by a grey node and $n-1$
white nodes (for $sl(1|2)$ this is the choice corresponding to case C),
see Eq.~(\ref{eq:distinguishedBorel}).  According to the dictionary,
we have adjoint fields $\Phi^a$ for every white node, and fundamentals
at each node. This means that for each white node, we can introduce a
superpotential of the type
\begin{align}
  W &= Q^a_{\phantom{a}k} \left( \Phi^a \right)^{\Lambda^{a}} \overline{Q^{a}}_k \, , & a &\neq m ,
\end{align}
which conserves a $U(1)$--symmetry, thus imposing a constraint on the
twisted masses:
\begin{align}
  \Lambda^{a} {{}\widetilde{m}^{\adj}}^{(a)} + {{}\widetilde{m}^{\fund}}^{(a)}_{k} + {{}\widetilde{m}^{\bar \fund}}^{(a)}_{k} &= 0  \, , & a &\neq m \, .
\end{align}
Requiring the superpotential to be a polynomial in the fields translates to the conditions
\begin{align}
  \Lambda^{a} &\in \setN \, ,\quad  a \neq m \, ;&  \Lambda^{m} &\in \setR \, .
\end{align}
This reproduces exactly the conditions that the representation
$\Lambda$ has to satisfy in order to be finite--dimensional (see
Appendix~\ref{sec:superalgebra}). In this case, the two sides of the
correspondence lead to the same constraints.


\section{Embedding in type IIA string theory}
\label{sec:brane-cartoons}

We have shown that the three quiver gauge theories introduced in
Section~\ref{sec:examples} have the same supersymmetric ground
states. It is reasonable to expect that this connection also manifests itself in other ways.
Here we show that they can also be related to each
other using a string theory embedding.  In this section, we propose a
possible mechanism based on brane transitions\footnote{We thank Kentaro Hori for suggesting this construction.} that
faithfully reproduces the matter content of our three quiver gauge
theories. In the present setup, the construction corresponds to 
vanishing twisted masses.  A complete type \textsc{iia} embedding that
reproduces the twisted masses and an M--theory description of the
transition are currently under investigation.

\bigskip

Brane constructions such as the ones
in~\cite{Hanany:1996ie,Hanany:1997vm,Brodie:1997wn} are likely
candidates for relating the three quiver gauge theories.  Our setup
(in type \textsc{iia} string theory) is the following. We consider two
parallel \NS5 branes $\NS5^{i},\ i=1,2$ which are extended in the
$012345$--directions, and another \NS5--brane $\NS5^{'}$ extended in
the $012389$--directions. There are
$N_a$ \D2--branes (extended in the $016$--directions) stretching between the \NS5--branes. Furthermore, we
have $L$ \D4--branes extended in the $ 01789$--directions. The setup
is summarized in Table~\ref{tab:setup}. This configuration preserves
$4$ of the $32$ supercharges of type \textsc{iia} string theory.  Note
the invariance under the rotations in the $(01)$, $(23)$, $(45)$ and
$(89)$ planes: these appear as Lorentz invariance and as global
symmetries in the field theory.

\begin{table}[h]\label{tab:setup}
  \centering
  \begin{tabular}{rcccccccccc}
    \toprule
    & 0 & 1 & 2 & 3 & 4 & 5 & 6 & 7 & 8 & 9 \\
    \midrule
    $\NS5^{1,2}$&$\times$&$\times$&$\times$ &$\times$ & $\times$& $\times$&& & &\\
    $\NS5^{'}$ &$\times$&$\times$&$\times$&$\times$&&&&&$\times$&$\times$ \\
    \D2 &$\times$&$\times$& & & & &$\times$& & & \\
    \D4 &$\times$&$\times$& & & &  & &$\times$&$\times$&$\times$ \\
    \bottomrule
  \end{tabular}
  \caption{Brane setup for the type \textsc{iia} embedding.}
\end{table}

Open strings stretching between \D2--branes located between parallel
\NS5--branes correspond to the adjoint fields in the $\mathcal{N}=2$
sector, while open strings stretching between two separate stacks of
\D2--branes correspond to bifundamental fields.  Open strings
stretching between the \D2 branes and the stack of $L$ \D4--branes
correspond to the fundamental and anti--fundamental fields of the
flavor group $U(L)$. The parameters of the field theory are encoded
in the positions of the \NS5 and \D4 branes. Here we restrict
ourselves to the case of all matter fields being
massless\footnote{Similar setups have been discussed
  in~\cite{Hanany:1997vm,Brodie:1997wn}.}.

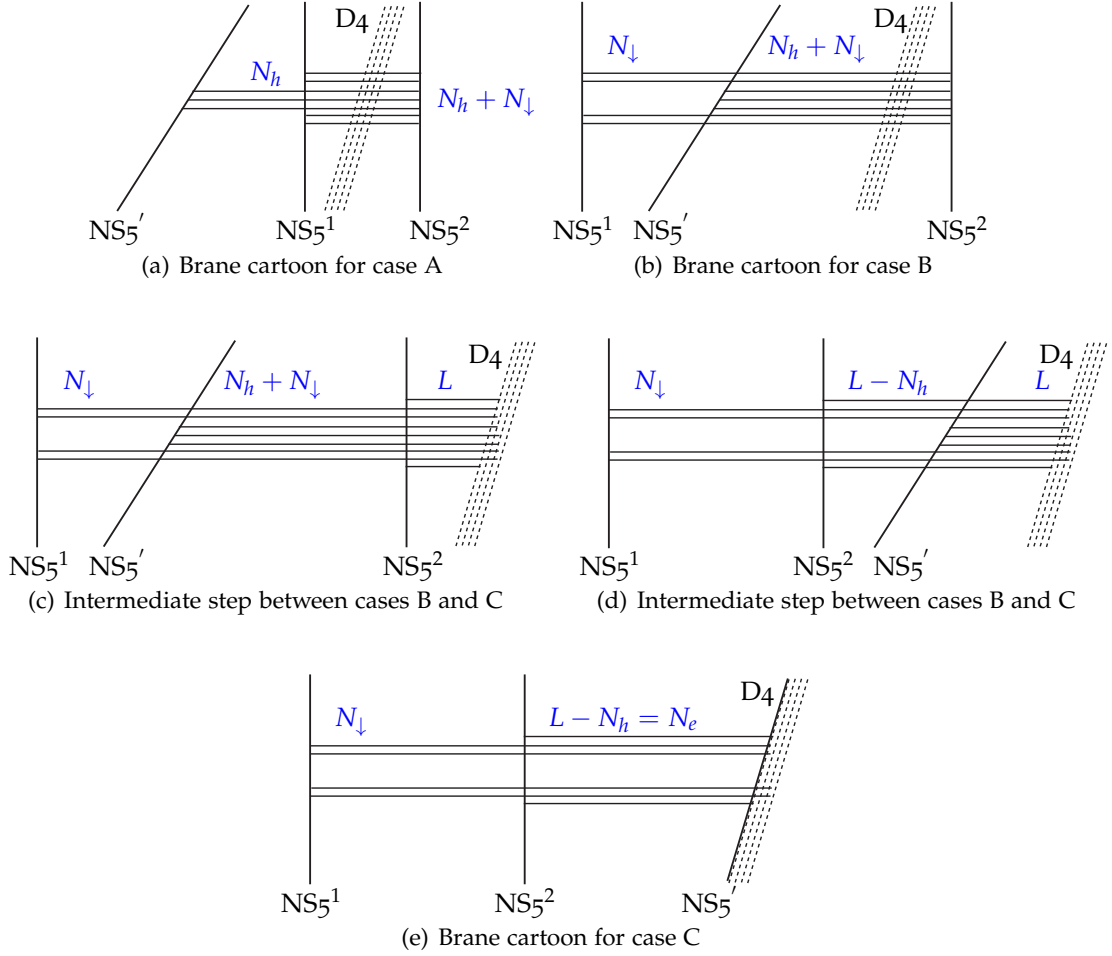
\begin{figure}
  \centering
  \subfigure[Brane cartoon for case A]{\label{fig:cart1}
      \begin{picture}(150,80)(,)
        \begin{small}
          \put(0,0){$\NS5^{'}$} \put(70,0){$\NS5^1$} \put(120,0){$\NS5^2$} \put(92,80){$\D4$}
          \textcolor{blue}{\put(60,60){$N_h$} \put(130,50){$N_h+N_\da$}}
        \end{small}
        \put(10,10){\includegraphics[scale=.7]{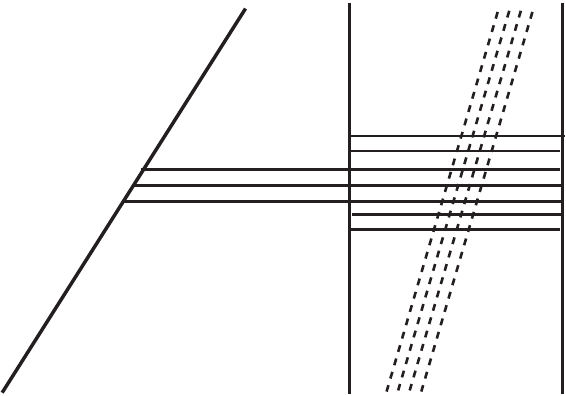}}
      \end{picture} }
    \hspace{1em}
    \subfigure[Brane cartoon for case B]{\label{fig:cart2}
      \begin{picture}(170,80)(,)
        \begin{small}
          \put(0,0){$\NS5^1$} \put(30,0){$\NS5^{'}$} \put(140,0){$\NS5^2$} \put(120,80){$\D4$}
          \textcolor{blue}{\put(20,70){$N_\da$} \put(80,70){$N_h+N_\da$}}
        \end{small}
        \put(10,10){\includegraphics[scale=.7]{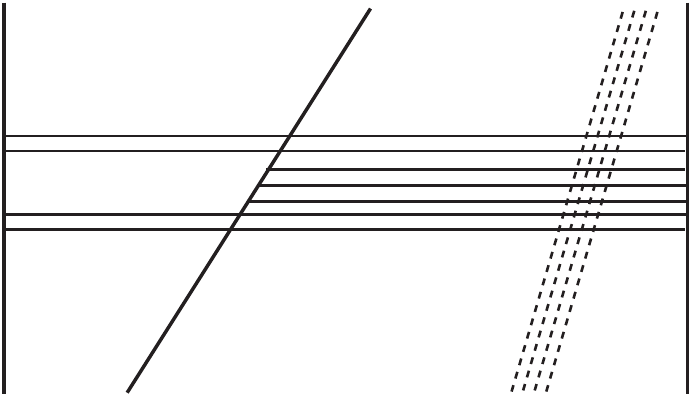}}
      \end{picture} }
    \hspace{1em}
    \subfigure[Intermediate step between cases B and C]{\label{fig:cart3}
      \begin{picture}(190,100)(,)
        \begin{small}
          \put(0,0){$\NS5^1$} \put(30,0){$\NS5^{'}$} \put(140,0){$\NS5^2$} \put(172,80){$\D4$}
          \textcolor{blue}{\put(20,70){$N_\da$} \put(80,70){$N_h+N_\da$} \put(160,70){$L$}}
        \end{small}
        \put(10,10){\includegraphics[scale=.7]{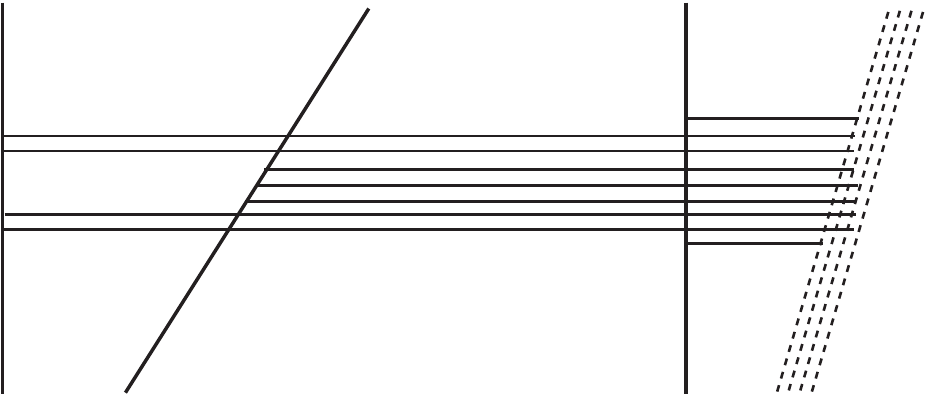}}
      \end{picture} }
    \hspace{1em}
    \subfigure[Intermediate step between cases B and C]{\label{fig:cart4}
      \begin{picture}(190,100)(,)
        \begin{small}
          \put(0,0){$\NS5^1$} \put(110,0){$\NS5^{'}$} \put(80,0){$\NS5^2$} \put(172,80){$\D4$}
          \textcolor{blue}{\put(20,70){$N_\da$} \put(100,70){$L-N_h$} \put(170,70){$L$}}
        \end{small}
        \put(10,10){\includegraphics[scale=.7]{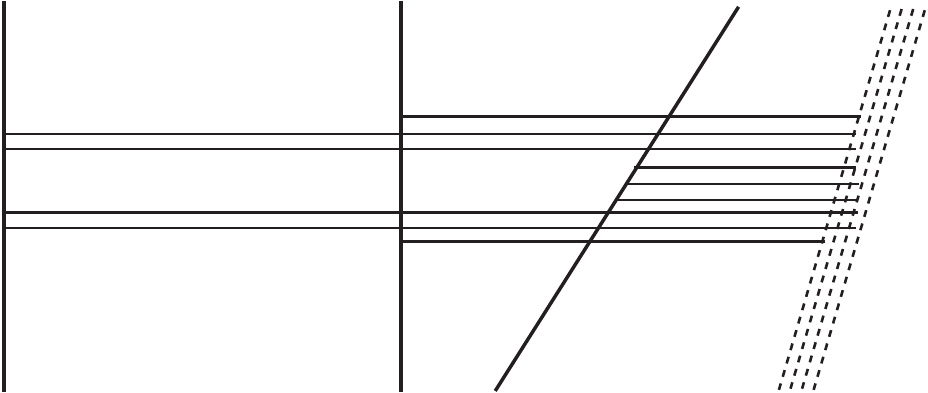}}
      \end{picture} }
    \hspace{1em}
    \subfigure[Brane cartoon for case C]{\label{fig:cart5}
      \begin{picture}(200,100)(,)
        \begin{small}
          \put(0,0){$\NS5^1$} \put(150,0){$\NS5^{'}$} \put(80,0){$\NS5^2$} \put(172,80){$\D4$}
          \textcolor{blue}{\put(20,70){$N_\da$} \put(100,70){$L-N_h = N_e$}}
        \end{small}
        \put(10,10){\includegraphics[scale=.7]{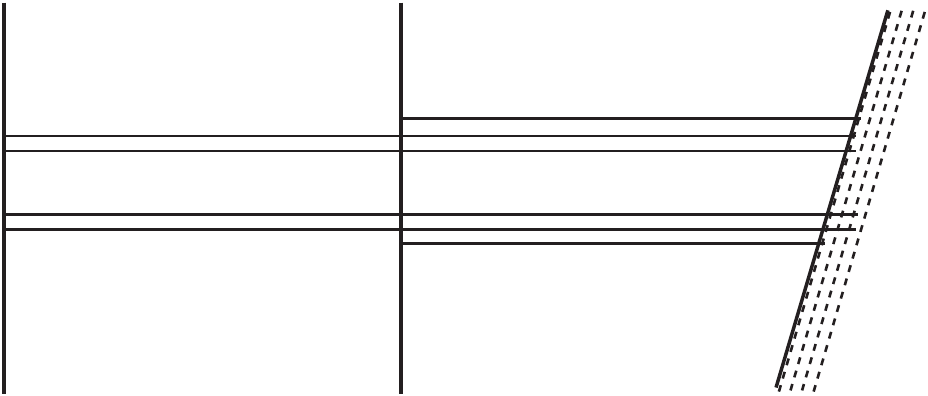}}
      \end{picture} }    
  \caption{Brane transitions connecting the quiver gauge theories of cases A, B, C}
  \label{fig:brane-cartoons}
\end{figure}

The transitions between the three different brane configurations work as follows:
\begin{enumerate}
\item We start with the brane cartoon in Figure~\ref{fig:cart1},
  corresponding to the quiver of case A
  (Figure~\ref{fig:quiver-A}). There are $N_h$ \D2 branes between
  $\NS5^{'}$ and $\NS5^1$, $N_h + N_\da$ \D2 branes between $\NS5^1 $ and
  $\NS5^2$, and the \D4 branes are located between the two parallel \NS5
  branes.
\item Moving the $\NS5^{'}$ brane past $\NS5^1$, we obtain the brane
  cartoon in Figure~\ref{fig:cart2} which corresponds to the setup of
  case B (Figure~\ref{fig:quiver-B}). Note that now, there are no \D2
  branes stretching between parallel \NS5 branes, which explains why
  there is no adjoint matter.
\item To reach the last configuration of case C
  (figure~\ref{fig:quiver-C}), we need to go through two intermediate
  steps. First, we move the $L$ \D4--branes to the right past
  $\NS5^2$, thus generating $L$ \D2 branes (see
  Figure~\ref{fig:cart3}).
\item Then we move also the $\NS5^{'}$--brane to
  the right, past $\NS5^2$ (see Figure~\ref{fig:cart4}). This creates
  an $\mathcal{N}=2$ sector on the left hand side of the cartoon (with
  the corresponding adjoint field).
\item In the last step, we make the $\NS5^{'}$--brane coincide with
  the \D4--branes. The resulting brane cartoon
  (Figure~\ref{fig:cart5}) corresponds to the quiver diagram of case C.
\end{enumerate}

With this, the three theories have been related by a mechanism completely different from the Gauge/Bethe correspondence, which can also be used as a further alley of investigation.

\section{Conclusions}
\label{sec:conclusions}

In this note, we have used the Gauge/Bethe correspondence by Nekrasov
and Shatashvili to relate 
\changed different supersymmetric quiver gauge
theories in two dimensions. These theories, despite having different
gauge groups and matter content, turn out to have the same chiral ring
and therefore the same supersymmetric ground states. We have thus used
quantum integrable systems as a tool to make statements about gauge
theories:
\begin{itemize}
\item \changed in the \textsc{xxx} case, we showed that a theory with $N$
  colors and $L$ flavors has the same supersymmetric ground states as a theory with
  $L-N$ colors and $L$ flavors;
\item in the $tJ$ model case, we used in particular the fact that
  integrable systems with supergroup symmetry give rise to several
  sets of Bethe equations, which correspond to different quiver gauge
  theories. In particular, in the $sl(m|n)$ case, there are
  $\binom{m+n}{m}$ equivalent quiver gauge theories with $m+n-1$
  nodes.
\end{itemize}
It is little surprising that the  gauge theories under
consideration can also be related via a string theory construction
using brane movements.

While the translation of two--dimensional supersymmetric gauge
theories into integrable systems is less straight--forward than going
in the opposite direction, we suggest to follow this path in oder to
gain knowledge about gauge theories via quantum integrable
systems.  The parameters of the quantum integrable models translate
into precise values for the twisted masses of the supersymmetric gauge
theories, which can be rather restrictive for the allowed values of
twisted masses. For integrable systems with supergroup symmetry, the
range of possible values is quite large, though: in the case of the
distinguished Borel subalgebra, one finds that nodes carrying adjoint
fields (white nodes) admit any non--negative integer weights, and
the weights for nodes without adjoint (grey nodes) are even continuous
parameters.  While the explicit \changed examples we used were based
on the \textsc{xxx} and  $tJ$
models, 
we believe
that our approach can be applied in a wider context. It is conceivable
to tune the twisted masses of the gauge theories under consideration
to values compatible with a spin chain embedding to check whether their
supersymmetric ground states can be matched. Once this relation is
established, other means of investigation such as the realization of
the systems via brane cartoons can be used to study the gauge theories
at different values for the twisted masses.


\subsection*{Acknowledgements}

We would like to thank Samson Shatashvili for introducing us to his
work which has inspired this note. We are deeply indebted to Kentaro
Hori for sharing his extensive knowledge of supersymmetric gauge
theories in two dimensions, as well as for his patience. We would
furthermore like to thank Simeon Hellerman for discussions and Nicolai
Reshetikhin for correspondence. Lastly, we would like to thank the referee for suggesting improvements. The research of the authors was
supported by the World Premier International Research Center
Initiative (WPI Initiative), MEXT, Japan.

\appendix

\section{The superalgebra $sl(1|2)$}
\label{sec:superalgebra}

In this appendix, we collect some facts about Lie superalgebras. For
details see~\cite{Kac:1977qb,Kac:1977em}.

\paragraph{Superalgebra and spin operators.}

A superalgebra $\mathfrak{g}$ can be decomposed into an even and an
odd part, $\mathfrak{g} = \mathfrak{g}_0 \oplus \mathfrak{g}_1 $. The
even part $\mathfrak{g}_0 = gl(1) \oplus sl(2) $ of the superalgebra
$sl(1|2)$ is generated by the operators $S^\pm, S^z, Z $ with
commutation relations
\begin{align}
  [ S^Z, S^\pm ] = \pm S^\pm \, , && [S^+, S^-] = 2 S^z \, , && [ Z, S^\pm ] = 0 \, , && [Z, S^z ] = 0 \, .
\end{align}
There are two additional fermionic multiplets $Q_s^\pm$, $s =
\set{\ua, \da} $ which transform as $(\pm \frac{1}{2}, \frac{1}{2})$
with respect to $\mathfrak{g}_0$. Explicitly:
\begin{align}
  [ S^z, Q_s^\pm ] &= \pm \frac{1}{2} Q_s \,, & [S^\pm, Q^\pm_s ] &= 0 \,, &
  [Z, Q^{\pm}_\da ] &= \frac{1}{2} Q^\pm_\da \, , & [Z, Q^\pm_\ua] &= - \frac{1}{2} Q^\pm_\ua \, .
\end{align}
The fermionic generators satisfy the following anticommutation relations:
\begin{align}
  \set{ Q^\pm_s, Q^\mp_s} = 0 \, , && \set{Q^\pm_\ua, Q^\pm_\da} = S^\pm \, , && \set{Q^\pm_\ua, Q^\mp_\da} = Z \pm S^z \, .  
\end{align}

The $tJ$ model admits a natural representation of the $sl(1|2) $
algebra. At each point $k$ of the lattice, the generators can be
represented in terms of creation--annihilation operators as
\begin{align}
  S^-_k &= c^\dagger_{k,\ua} c_{k,\da} \, , & S_k^+ &= c^\dagger_{k,\da} c_{k,\ua} \, , & S_k^z &= \frac{1}{2} \left( n_{k,\ua} - n_{k,\da} \right) \, ,\\
  Q^-_{k,\ua} &= \left( 1 - n_{k,\da} \right) c_{k, \ua} \, , &
  Q^+_{k,\ua} &= \left( 1 - n_{k,\da} \right) c^{\dagger}_{k,\ua} \, ,
  & Q^-_{k,\da} &= \left( 1 - n_{k,\ua} \right) c_{k, \da} \, , \\
  Q^+_{k,\da} &= \left( 1 - n_{k,\ua} \right) c^{\dagger}_{k,\da}
  \, , & Z_k &= 1 - \frac{1}{2} n_k \, .
\end{align}

\paragraph{Root decomposition.}

Let $\set{\delta_1, \dots, \delta_m, \epsilon_1, \dots, \epsilon_n}$
be a basis for $\setC^{(m|n)}$ with inner product
\begin{align}
  ( \delta_i, \delta_j ) &= \delta_{ij} \, , & ( \epsilon_{\bar{i}}, \epsilon_{\bar j}  )
  &= - \delta_{\bar{i}\bar{j}} \, ,& (\delta_i, \epsilon_{\bar{i}} ) &=0 \, .
\end{align}
The superalgebra $sl(m|n)$ admits a root space decomposition
\begin{equation}
  sl(m|n) = \mathfrak{h} \oplus \bigoplus_{\alpha \in \Phi} \mathfrak{g}_\alpha \, ,  
\end{equation}
where $\mathfrak{h}$ is the Cartan subalgebra (diagonal matrices) and
$\Phi$ is the root system:
\begin{equation}
  \Phi = \set{\delta_i - \delta_j} \cup \set{\epsilon_{\bar i}- \epsilon_{\bar j} } \cup \set{\pm \left( \delta_i - \epsilon_{\bar i} \right)} \hspace{2em} i,j = 1, \dots, m \, ; \, \bar i, \bar j = 1, \dots, n \, .  
\end{equation}
The \emph{standard set} of positive roots (corresponding to the
\emph{distinguished Borel subalgebra}) is
\begin{equation}
  \Phi^+ = \set{\delta_i - \delta_j| i < j } \cup \set{\epsilon_{\bar i} - \epsilon_{\bar j}| \bar i < \bar j } \cup \set{ \delta_i - \epsilon_{\bar i} }  \hspace{2em} i,j = 1, \dots, m \, ; \, \bar i, \bar j = 1, \dots, n \, ,
\end{equation}
where $\set{\delta_i - \epsilon_{\bar i} }$ are odd roots, \emph{i.e.}
\begin{equation}
  ( \delta_i - \epsilon_{\bar i}, \delta_i - \epsilon_{\bar i} ) = 0 \, .  
\end{equation}

For Lie algebras, all possible sets of positive roots can be obtained
by reflection, and the corresponding Borel subalgebras $\mathfrak{b} =
\mathfrak{h} + \mathfrak{n}^+$ are conjugate.  This is not the case
for superalgebras, since the reflection of an odd root $\delta_i -
\epsilon_{\bar i} \to \epsilon_{\bar i} - \delta_i$ produces a new
system of positive roots whose associated Borel subalgebra is not
conjugate to the initial one. For $sl(1|2)$, there are six possible
choices of positive roots, which are organized into three conjugacy
classes under reflection. It is convenient to represent the positive
roots by using the Cartan matrix or, equivalently, a Dynkin
diagram. Now we need to distinguish between even roots (white nodes
$\ocircle$) and odd roots (grey nodes $\otimes$). The standard Dynkin
diagrams and the other two obtained by reflection of odd roots are
represented in Table~\ref{tab:root-decompositions}.  In the general
$sl(m|n)$ case, there are  $\binom{m+n}{m}$ conjugacy classes (and Dynkin
diagrams), one for each sequence of $m$ repetitions of the symbol
$\delta$ and $n$ repetitions of the symbol $\epsilon$.
\begin{table}
  \centering
  \begin{tabular}{ccc}
    \toprule
    $\Phi^+$ & $C^{ab}$ & Dynkin diagram \\ \midrule
    $\set{\delta_1 -\epsilon_1, \epsilon_1 - \epsilon_2}$ & $
    \begin{pmatrix}
      0 & -1 \\ -1 & 2
    \end{pmatrix}$ & $\diagA{}{}$\\
    $\set{\epsilon_1 - \delta_1, \delta_1 - \epsilon_2}$ & $
    \begin{pmatrix}
      0 & -1 \\ -1 & 0
    \end{pmatrix}$ & $\diagB{}{}$ \\
    $\set{\epsilon_1 -\epsilon_2, \epsilon_2 - \delta_1}$ & $
    \begin{pmatrix}
      2 & -1 \\ -1 & 0
    \end{pmatrix}$ & $\diagC{}{}$ \\ \bottomrule
  \end{tabular}
  \caption{Non--equivalent root decompositions for $sl(1|2)$. Positive roots, Cartan matrix and Dynkin diagram.}
  \label{tab:root-decompositions}
\end{table}

\paragraph{Kac--Dynkin diagrams.}

Representations of $sl(m|n)$ are labelled uniquely by so-called {Kac--Dynkin}
diagrams. These are Dynkin diagrams in which a number
$\Lambda^{a}$ is associated to each node. For example, the fundamental
representation for $sl(1|2)$ can be associated to three
non--equivalent Kac--Dynkin diagrams:
\begin{align}
  \text{(case A) } & \diagA{0}{1} & \text{(case B) } &\diagB{0}{1} & \text{(case C) } &\diagC{0}{1}  \, .
\end{align}
If we choose the distinguished Borel subalgebra, a representation
\begin{equation}
\label{eq:distinguishedBorel}
  \Lambda = 
  \begin{scriptsize}
    \begin{tikzpicture}
      \node (a) at (0,0) [circle, draw, inner sep=2pt, minimum size=3mm, label=above:$\Lambda^{1}$] {}; 
      \node (b) at (1,0) [circle, draw, inner sep=2pt, minimum size=3mm, label=above:$\Lambda^{2}$] {};
      \node (c) at (3,0) [circle, draw, inner sep=2pt, minimum size=3mm, label=above:$\Lambda^{m-1}$] {};
      \node (d) at (4,0) [crossed circle, draw, inner sep=2pt, minimum size=3mm, label=above:$\Lambda^{m}$] {};
      \node (e) at (5,0) [circle, draw, inner sep=2pt, minimum size=3mm, label=above:$\Lambda^{m+1}$] {};
      \node (f) at (7,0) [circle, draw, inner sep=2pt, minimum size=3mm, label=above:$\Lambda^{m+n-1}$] {};
      \draw (a) to (b);
      \draw (b) to (1.5,0);
      \draw[dashed] (1.5,0) to (2.3,0);
      \draw (2.3,0) to (c);
      \draw (c) to (d);
      \draw (d) to (e);
      \draw (e) to (5.5,0);
      \draw[dashed] (5.5,0) to (6.3,0);
      \draw (6.3,0) to (f);
    \end{tikzpicture}
  \end{scriptsize}
\end{equation}
is finite dimensional if and only if the labels of the white
nodes are non--negative integers and the label of the grey node
$\Lambda^{m}$ is a real number.

\section{Equivalence of $tJ$ Bethe Equations}
\label{sec:equivalence}

We want to show the equivalence of the three Bethe ansatz equations
described in Section~\ref{sec:example:-tj-model} explicitly by using an
argument originally introduced in~\cite{PhysRevB.46.14624}.

Consider the Bethe ansatz equations in Eq.~\eqref{eq:Bethe-B}:
\begin{subequations}
 \begin{align}
    \left(\frac{\lambda^{(2)}_p + \frac{i}{2}}{\lambda^{(2)}_p - \frac{i}{2}}\right)^L&= \prod_{i=1}^{N_{\da}}\frac{\lambda^{(1)}_i - \lambda^{(2)}_p - \frac{i}{2}}{\lambda^{(1)}_i - \lambda^{(2)}_p + \frac{i}{2}} \, , \quad p = 1,\dots, N_h+N_\da \\
    1 &= \prod_{p=1}^{N_h+N_{\da}}\frac{\lambda_i^{(1)} -
      \lambda^{(2)}_p - \frac{i}{2}}{\lambda_i^{(1)}-\lambda^{(2)}_p +
      \frac{i}{2}}\, , \quad i=1,\dots,N_{\da} \, .
  \end{align}
\end{subequations}
The second set (the unknowns are $\lambda_i^{(1)}$) can be written as
a polynomial equation:
\begin{equation}
  1 = \prod_{p=1}^{N_h+N_{\da}}\frac{\lambda_i^{(1)} -
    \lambda^{(2)}_p - \frac{i}{2}}{\lambda_i^{(1)} - \lambda^{(2)}_p +
    \frac{i}{2}}  \Leftrightarrow p(w) = \prod_{p=1}^{N_h+N_{\da}} \left( w -
  \lambda^{(2)}_p - \frac{i}{2}\right) -  \prod_{p=1}^{N_h+N_{\da}} \left( w-\lambda^{(2)}_p +
    \frac{i}{2} \right)  = 0\, .
\end{equation}
The polynomial $p(w)$ has degree $N_\da + N_h$. We identify the
variables $\lambda_i^{(1)} $ with the first $N_\da$ solutions $w_i$ of
$p(w) = 0$. We call the other $N_h$ solutions $\bar w_j $, $j=1,
\dots, N_h$.

\begin{figure}
  \centering
  \subfigure[Integration contour in Eq.~\eqref{eq:integration-contour-A}]{\label{fig:integration-1}
     \setlength\fboxsep{0pt}
     \setlength\fboxrule{0.5pt}
     \fbox{
       \begin{picture}(185,150)(,)
         \begin{small}
          \put(0,100){$w_1$} \put(40,100){$w_2$} \put(90,97){$w_{N_\da}$}
          \put(90,130){$\bar w_1$} \put(115,120){$\bar w_2$} \put(170,95){$\bar w_{N_h}$}
          \put(150,20){$\lambda^{(2)}_p$}
        \end{small}
        \put(10,10){\includegraphics[scale=1]{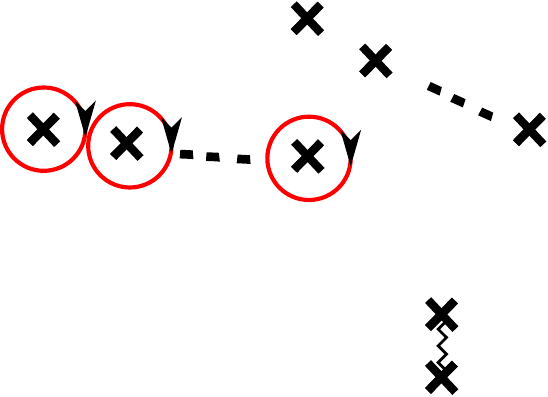}}
      \end{picture} }}
    \hspace{1em}
    \subfigure[Integration contour in Eq.~\eqref{eq:integration-contour-B}]{\label{fig:integration-2}
      \fbox{\begin{picture}(180,150)(,)
        \begin{small}
          \put(0,100){$w_1$} \put(40,97){$w_2$} \put(90,90){$w_{N_\da}$}
          \put(92,136){$\bar w_1$} \put(120,120){$\bar w_2$} \put(170,95){$\bar w_{N_h}$}
           \put(140,20){$\lambda^{(2)}_{p}$}
        \end{small}
        \put(10,10){\includegraphics[scale=1]{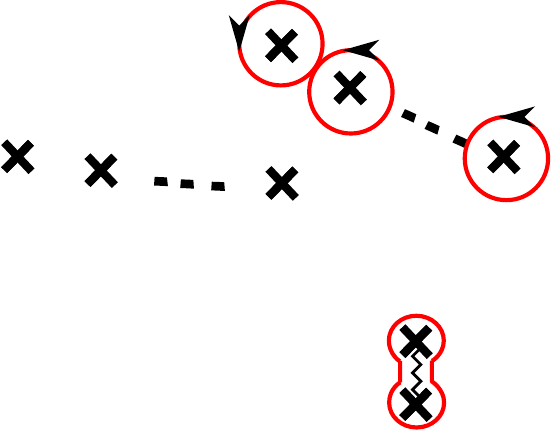}}
      \end{picture} } }
  \caption{Equivalent integration contours around the poles of $p(z)$.}
  \label{fig:Integration-Contours}
\end{figure}
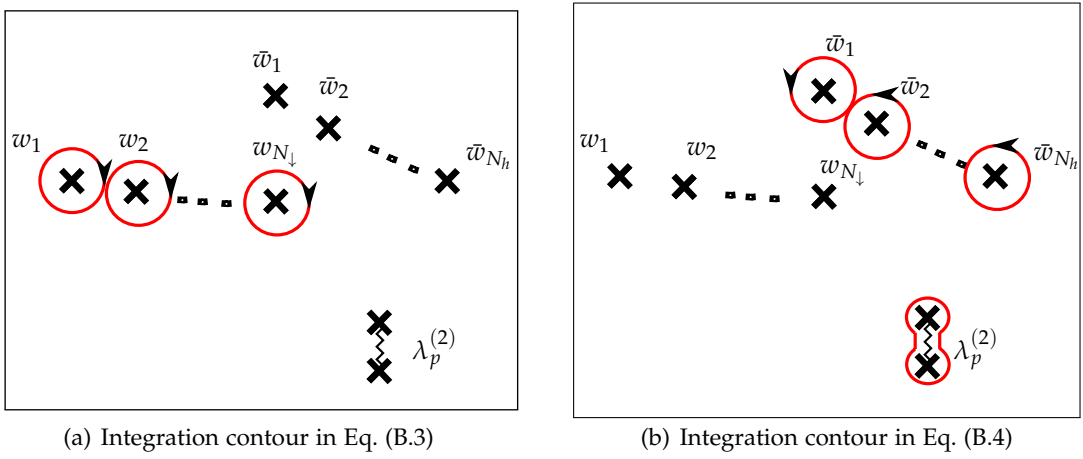

Using the residue theorem, we can write the \textsc{rhs} of the first
set of BEA (Eq.~\eqref{eq:Bethe-B}(a)) as
\begin{equation}
  \label{eq:integration-contour-A}
  \sum_{i=1}^{N_{\da}} \log \left[  \frac{ \lambda^{(2)}_p - w_i + \frac{i}{2}}{ \lambda^{(2)}_p - w_i - \frac{i}{2}} \right] = \sum_{i=1}^{N_\da} \frac{1}{2 \pi \imath} \oint_{C_i} dz \, \log \left[ \frac{ \lambda^{(2)}_p - z + \frac{i}{2}}{ \lambda^{(2)}_p - z - \frac{i}{2}} \right] \frac{d}{dz} \log ( p(z) ) \, ,
\end{equation}
where $C_i$ is a contour around $w_i$ (see
Figure~\ref{fig:integration-1}). The logarithm has a branch cut from
$(\lambda^{(2)}_p + \imath/2) $ to $ (\lambda^{(2)}_p - \imath /2)
$. We can change the contour, picking residues from the other $N_h$
poles of $p(z)$, plus the contributions of the branch cut (see
Figure~\ref{fig:integration-2}):
\begin{multline}
  \label{eq:integration-contour-B}
  \sum_{i=1}^{N_{\da}} \log \left[  \frac{ \lambda^{(2)}_p - w_i + \frac{i}{2}}{ \lambda^{(2)}_p - w_i - \frac{i}{2}} \right] \\= -\sum_{j=1}^{N_h} \frac{1}{2 \pi \imath} \oint_{C_j} \di z \, \log \left[ \frac{ \lambda^{(2)}_p - z + \frac{i}{2}}{ \lambda^{(2)}_p - z - \frac{i}{2}} \right] \frac{d}{dz} \log ( p(z) ) - \log \left[ \frac{p(\lambda^{(2)}_p + \imath /2 ) }{p(\lambda^{(2)}_p - \imath /2 )} \right] \\ 
  = - \sum_{j=1}^{N_{h}} \log \left[  \frac{ \lambda^{(2)}_p - \tilde w_j + \frac{i}{2}}{ \lambda^{(2)}_p - \tilde w_j - \frac{i}{2}} \right] - \log \left[ \frac{p(\lambda^{(2)}_p + \imath /2 ) }{p(\lambda^{(2)}_p - \imath /2 )} \right] \, .
\end{multline}
Writing $p(z)$ explicitly:
\begin{align}
  p ( \lambda^{(2)}_p + \imath /2 ) = - \prod_{\substack{q=1\\q \neq p}}^{N_h+N_{\da}}
  \left( \lambda^{(2)}_p - \lambda^{(2)}_q + \imath \right) \, ,&& p ( \lambda^{(2)}_p - \imath /2 ) =  \prod_{\substack{q=1\\ q\neq p}}^{N_h+N_{\da}}
  \left( \lambda^{(2)}_p - \lambda^{(2)}_q - \imath \right) \, ,
\end{align}
we can exponentiate:
\begin{equation}
  \prod_{i=1}^{N_{\da}}   \frac{ \lambda^{(2)}_p - w_i + \frac{i}{2}}{ \lambda^{(2)}_p - w_i - \frac{i}{2}} = \prod_{j=1}^{N_{h}}  \frac{ \lambda^{(2)}_p - \tilde w_j - \frac{i}{2}}{ \lambda^{(2)}_p - \tilde w_j + \frac{i}{2}} \prod_{\substack{q=1\\q \neq p}}^{N_h+N_\da} \frac{\lambda^{(2)}_p - \lambda^{(2)}_q + \imath}{\lambda^{(2)}_p - \lambda^{(2)}_q - \imath} \, ,
\end{equation}
and using the set in Eq.~(\ref{eq:Bethe-B}a) on the \textsc{lhs}, we
find
\begin{equation}
  \left(\frac{\lambda^{(2)}_p+\frac{i}{2}}{\lambda^{(2)}_p-\frac{i}{2}}\right)^L
  = \prod_{\substack{q=1\\ q\neq p}}^{N_h+N_{\da}}\frac{
    \lambda^{(2)}_p - \lambda^{(2)}_q + {i}}{\lambda^{(2)}_p -
    \lambda^{(2)}_q - {i}}\prod_{i=1}^{N_h} \frac{\lambda^{(2)}_p -
    \bar w_i - \frac{i}{2}}{\lambda^{(2)}_p - \bar w_i +\frac{i}{2}},
  \quad p=1,\dots,N_h+N_{\da} \, ,
\end{equation}
which coincides with the equation of case A Eq.~(\ref{eq:Bethe-A}a) if
we identify $\bar w_i = \lambda^{(1)}_i$:
\begin{subequations}
 \begin{align}
    \left(\frac{\lambda^{(2)}_p+\frac{i}{2}}{\lambda^{(2)}_p-\frac{i}{2}}\right)^L
    &= \prod_{\substack{q=1\\ q\neq p}}^{N_h+N_{\da}}\frac{
      \lambda^{(2)}_p - \lambda^{(2)}_q + {i}}{\lambda^{(2)}_p -
      \lambda^{(2)}_q - {i}}\,\prod_{i=1}^{N_h} \frac{\lambda^{(2)}_p -
      \lambda_i^{(1)} - \frac{i}{2}}{\lambda^{(2)}_p - 
      \lambda_i^{(1)}+\frac{i}{2}},
    \quad p=1,\dots,N_h+N_{\da}, \\
    1 &=
    \prod_{p=1}^{N_h+N_{\da}}\frac{\lambda^{(2)}_p-\lambda_i^{(1)}-\frac{i}{2}}{\lambda^{(2)}_p
      - \lambda_i^{(1)}+\frac{i}{2}}, \quad i=1,\dots,N_{h}.
  \end{align}
\end{subequations}
The identification reproduces also the other equations
(\ref{eq:Bethe-A}b) since these can be put into the very same
polynomial form $p(w) = 0$, and the $\bar w_j $ are by construction
solutions.

\newpage

\bibliography{GaugeBetheRef}

\end{document}